\documentclass[12pt]{article}
\pdfoutput=1

\usepackage[a4paper,text={16.8cm,22.4cm}]{geometry}
\usepackage{amsmath,amsfonts,slashed,amssymb,tikz,bm,psfrag,graphicx,color,dsfont}
\usepackage{multicol}
\usepackage{float}

\RequirePackage[sort&compress,square,comma,numbers]{natbib}
\allowdisplaybreaks
\begin{document}

\begin{titlepage}

\begin{flushright}
\normalsize
 UWTHPH 2017-12 \\
June 18, 2017
\end{flushright}

\vspace{0.1cm}
\begin{center}
\Large\bf
Subleading power corrections to the pion-photon transition form factor in QCD
\end{center}

\vspace{0.5cm}
\begin{center}
{\bf Yu-Ming Wang$^{a,b}$,  Yue-Long Shen$^{c}$} \\
\vspace{0.7cm}
{\sl   ${}^a$ \, School of Physics, Nankai University, Weijin Road 94, 300071 Tianjin, China \\
 ${}^b$ \,  Fakult\"{a}t f\"{u}r Physik, Universit\"{a}t Wien, Boltzmanngasse 5, 1090 Vienna, Austria \\
 ${}^c$ \, College of Information Science and Engineering,
Ocean University of China, Songling Road 238, Qingdao, 266100 Shandong, P.R. China
}
\end{center}

\vspace{0.2cm}
\begin{abstract}

We reconsider QCD  factorization for the leading power contribution to the $\gamma^{\ast} \gamma \to \pi^0$
form factor $F_{\gamma^{\ast} \gamma \to \pi^0} (Q^2)$ at one loop using the evanescent operator approach,
and demonstrate the equivalence of  the resulting factorization formulae derived with distinct
prescriptions of $\gamma_5$ in dimensional regularization.
Applying the light-cone QCD sum rules (LCSRs) with photon distribution amplitudes (DAs)
we further compute the subleading power contribution to the pion-photon form factor induced by the
``hadronic" component of the real photon at the next-to-leading-order in ${\cal O}(\alpha_s)$, with both
naive dimensional regularization and 't Hooft-Veltman schemes of $\gamma_5$.
Confronting our theoretical predictions of $F_{\gamma^{\ast} \gamma \to \pi^0} (Q^2)$ with
the experimental measurements from the BaBar and the Belle Collaborations implies that a reasonable agreement
can be achieved without introducing an ``exotic" end-point behaviour for the twist-2 pion DA.

\end{abstract}

\vfil

\end{titlepage}

\section{Introduction}
\label{sect:Intro}

Hard exclusive processes play a prominent role  in exploring the strong interaction
dynamics of hadronic reactions in the framework of QCD.
The pion-photon  transition form factor  $\gamma^{\ast} \gamma \to \pi^{0}$ at large momentum transfers ($Q^2$)
serves as one of the simplest exclusive processes for testing the theoretical predictions
based upon perturbative QCD factorization. The hard-collinear factorization theorem for the pion-photon
form factor $F_{\gamma^{\ast} \gamma \to \pi^0} (Q^2)$ can be demonstrated at leading power in $1/Q^2$
utilizing both diagrammatic approaches \cite{Lepage:1980fj,Efremov:1979qk,Duncan:1979ny}
and effective field theory techniques \cite{Rothstein:2003wh}.
The hard coefficient function entering the leading-twist factorization formula has been computed
at one loop \cite{delAguila:1981nk,Braaten:1982yp,Kadantseva:1985kb},
and at two loops \cite{Melic:2002ij} in the large $\beta_0$ approximation.
In virtue of the fact that the twist-2 pion distribution amplitude (DA) is defined by
an axial-vector light-ray operator, a subtle issue in evaluating QCD corrections to the
hard function in dimensional regularization lies in the definition the Dirac matrix $\gamma_5$
in  the complex $D$-dimensional space demanding a new set of algebraic identities and various
prescriptions for the treatment of $\gamma_5$ have been proposed to meet the demand of precision QCD
calculations in different contexts (see \cite{Bonneau:1990xu,Collins:1984xc} for an overview
and \cite{Larin:1993tq,Martin:1999cc,Jegerlehner:2000dz,Moch:2015usa,Gutierrez-Reyes:2017glx} for more discussions).
Employing the trace technique, the $\gamma_5$ ambiguity of dimensional regularization
was resolved  by adjusting the way of manipulating $\gamma_5$ in each diagram
to preserve the axial-vector Ward identity  \cite{Braaten:1982yp}, which is less straightforward (systematic)
for the higher-order QCD calculations of hadronic reactions.
One of our major objectives of this paper is to demonstrate the equivalence of  factorization formulae
for the pion-photon transition form factor constructed with naive dimensional regularization (NDR)
and 't Hooft-Veltman (HV) schemes of the $\gamma_5$ matrix, using the spinor decomposition  technique
\cite{Beneke:2004rc,Beneke:2005gs,Beneke:2005vv} and the evanescent operator approach \cite{Dugan:1990df,Herrlich:1994kh}.

Confronting the theoretical predictions with the precision experimental measurements of the
$\pi^0 \gamma^{\ast} \gamma$ form factor at accessible $Q^2$ evidently necessities
a better understanding of the subleading power terms in the large momentum expansion,
due in particular to the scaling violation implied by the BaBar data  \cite{Aubert:2009mc}.
The significance of the power suppressed  contributions to $F_{\gamma^{\ast} \gamma \to \pi^0} (Q^2)$
was highlighted by evaluating the soft correction to the leading twist effect with
the dispersion approach \cite{Agaev:2010aq,Agaev:2012tm}  and turned out to be crucial to suppress the contributions
from higher Gegenbauer moments of the twist-2 pion DA (see also \cite{Kroll:2010bf,Li:2013xna}).
An attractive advantage of the dispersion approach \cite{Khodjamirian:1997tk} is that
the subleading power ``hadronic" photon correction is taken into account effectively  by  modifying the spectral function
in the real-photon channel at the price of introducing two nonperturbative parameters
(i.e., the vector meson mass $m_{\rho}$ and the  effective threshold parameter $s_0$).
This effective method allows  continuous improvement of the theoretical accuracy for predicting
the pion-photon form factor by including the next-to-next-to-leading order (NNLO) QCD correction
to the twist-2 contribution and the finite-width effect of the unstable vector mesons in the hadronic
dispersion relation \cite{Stefanis:2012yw,Bakulev:2011rp,Bakulev:2012nh,Mikhailov:2016klg}.
Further applications of this technique were  pursued in radiative leptonic $B$-meson decay
\cite{Braun:2012kp,Wang:2016qii} and electro-production
of  the pseudoscalar eta mesons \cite{Agaev:2014wna} and of  tensor mesons \cite{Braun:2016tsk}
in an attempt to ``overcome" the difficulty of rapidity divergences emerged in  the direct QCD calculations
of the subleading power contributions. It is then in demand to provide an independent QCD approach to  compute
the above-mentioned  power corrections for the sake of boosting our confidence on the reliability of both theoretical tools.
Another objective of this paper is to construct the light-cone sum rules (LCSRs) for the hadronic photon effect in the
pion-photon transition form factor with photon distribution amplitudes (DAs) \cite{Ball:2002ps}
at next-to-leading order (NLO) in $\alpha_s$.

Applying the transverse-momentum-dependent (TMD) factorization scheme for hard exclusive processes \cite{Li:1992nu},
the leading power contribution to the pion-photon form factor was also computed at ${\cal O}(\alpha_s)$
with the diagrammatic approach \cite{Nandi:2007qx} (see also \cite{Musatov:1997pu,Wu:2010zc}),
and the joint summation of the parametrically  large logarithms $\ln^2{{\bf k_{\perp}^2} /Q^2}$
and $\ln^2 x$ in the hard matching coefficient was performed in Mellin and impact-parameter spaces \cite{Li:2013xna}.
However, the subleading power contribution to $F_{\gamma^{\ast} \gamma \to \pi^0} (Q^2)$ has not been discussed systematically
in TMD factorization (see however \cite{Chen:2011pn} in the context of the pion electromagnetic form factor).
Further development of the TMD factorization for the $\pi^0 \gamma^{\ast} \gamma$ form factor with a definite power counting scheme
for the intrinsic transverse momentum and of the factorization-compatible TMD pion wave functions \cite{Li:2014xda}
will be essential to put this factorization scheme on a solid ground, albeit with the intensive applications
to many hard exclusive processes \cite{He:2006ud,Lu:2009cm,Li:2010nn,Li:2012nk,Li:2012md}.
The dedicated BaBar and Belle measurements  \cite{Aubert:2009mc,Uehara:2012ag} of $F_{\gamma^{\ast} \gamma \to \pi^0} (Q^2)$
also stimulated intensive theoretical investigations with various phenomenological approaches as well as
lattice QCD  simulations (see for instance \cite{Masjuan:2012wy,Hoferichter:2014vra,Gerardin:2016cqj}).
In particular, an ``exotic" twist-two pion DA with the non-vanishing end-point behaviour was proposed
\cite{Radyushkin:2009zg,Polyakov:2009je} to accommodate the  anomalous  BaBar data at high $Q^2$,
but was soon critically examined in \cite{Agaev:2010aq} concluding that a reasonable description of the BaBar data
in \cite{Radyushkin:2009zg,Polyakov:2009je} is achieved rather due to the introduction of  a sizable nonperturbative
soft correction from the TMD pion wavefunction.

The outline of this paper is  as follows:  in Section \ref{sect:leading-power effect} we recalculate
the one-loop hard  function entering the factorization formula for the pion-photon form factor
at leading power in $1/Q^2$ with both the NDR and HV schemes of $\gamma_5$, and demonstrate the renormalization-scheme
independence of the factorization formulae for physical quantities explicitly. It will be shown that our expression of the NLO
hard-scattering kernel in the NDR scheme reproduces the classical result obtained by Braaten \cite{Braaten:1982yp}
and the renormalization-prescription  dependence of the short-distance coefficient at ${\cal O}(\alpha_s)$
will be cancelled precisely by the scheme dependent twist-2 pion DA
at one loop. We then establish QCD factorization for the vacuum-to-photon correlation function defined with
a pseudoscalar interpolating current for the pion state and an electromagnetic current
carrying a space-like momentum $q_{\mu}$ ($q^2=-Q^2$) at one loop in Section  \ref{sect:subleading-power effect}.
It will be also  proved manifestly that the resulting hard matching coefficients
obtained in the NDR and HV schemes are related by the finite renormalization constant term, which is
 introduced in the HV scheme in order to fulfill the Adler-Bardeen theorem for the non-renormalization
of the axial anomaly \cite{Larin:1993tq,Matiounine:1998re,Ravindran:2003gi}.
The next-to-leading-logarithmic (NLL) resummation improved LCSR for the hadronic photon correction to
$F_{\gamma^{\ast} \gamma \to \pi^0} (Q^2)$ will be further presented  with the aid of the parton-hadron duality ansatz.
Taking advantage of the newly derived subleading power correction and the twist-four effect from
both the two-particle and three-particle pion DAs at tree level \cite{Khodjamirian:1997tk,Agaev:2010aq},
we will provide updated theoretical predictions for the pion-photon form factor in Section \ref{sect:numerical analysis}
with distinct nonperturbative models for the twist-2 pion DA.
A summary of our observations and the concluding remarks are presented in Section \ref{sect:conclusion}.
We collect the two-loop evolution functions for the leading twist DAs of the pion and the photon
in Appendix \ref{app:two-loop evolutions}    and display the spectral representations of the convolution integrals
for the construction of the NLL LCSR of the hadronic photon contribution in Appendix \ref{app:spectral representations}.

\section{Factorization of the leading power contribution}
\label{sect:leading-power effect}

The purpose of this section is to compute the leading power  contribution to the pion-photon
form factor at one loop
\begin{eqnarray}
\langle \pi(p) | j_{\mu}^{\rm em}  | \gamma (p^{\prime}) \rangle
= g_{\rm em}^2 \, \epsilon_{\mu \nu \alpha \beta} \, q^{\alpha} \, p^{\beta} \, \epsilon^{\nu}(p^{\prime})
F_{\gamma^{\ast} \gamma \to \pi^0} (Q^2) \,,
\end{eqnarray}
with  both the NDR and HV schemes for the $\gamma_5$ matrix in $D$ dimensions,
where $q=p-p^{\prime}$, $p$ refers to the four-momentum of the pion,
the on-shell photon carries the four-momentum $p^{\prime}$ and
\begin{eqnarray}
j_{\mu}^{\rm em} = \sum_q \, g_{\rm em} \, Q_q \, \bar q \, \gamma_{\mu} \, q \,, \qquad
\epsilon_{0123}= -1 \,.
\label{em current definition}
\end{eqnarray}
We further introduce a light-cone vector $\bar n_{\mu}$ parallel to the photon momentum $p^{\prime}$,
define another light-cone vector $n_{\mu}$ along the direction of the momentum $p$
in the massless pion limit, and employ the following power counting scheme at large momentum transfer
\begin{eqnarray}
\bar n \cdot p  \sim n \cdot p^{\prime} \sim {\cal O}(\sqrt{Q^2}) \,,
 \qquad  n \cdot p  \sim {\cal O}(\Lambda^2 / \sqrt{Q^2})  \,.
\end{eqnarray}

\begin{figure}
\begin{center}
\includegraphics[width=0.80 \columnwidth]{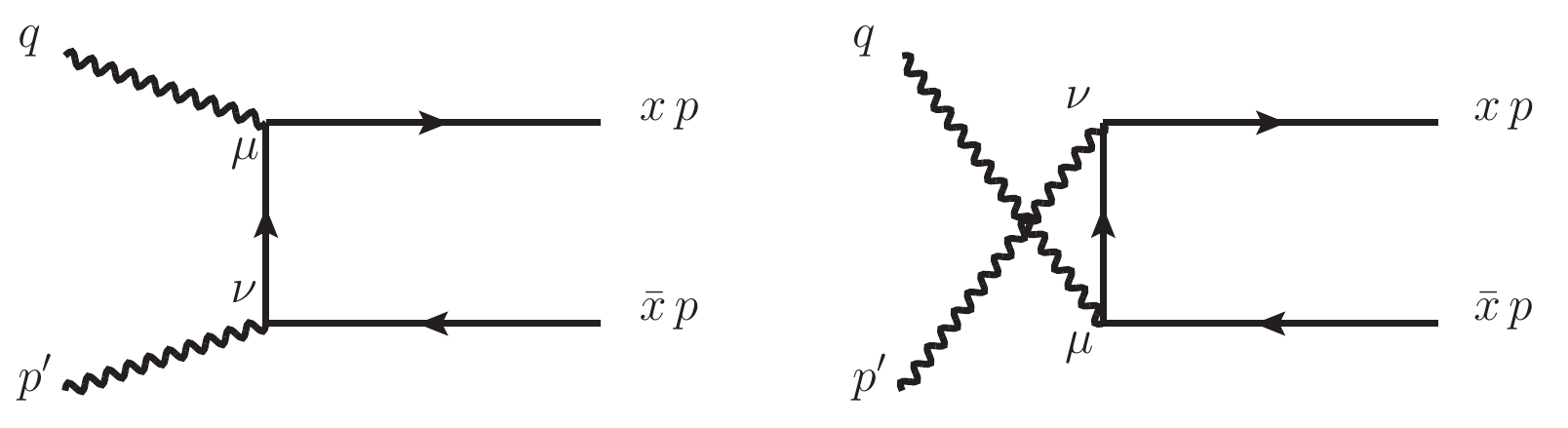} \\
(a) \hspace{6 cm} (b)
\vspace*{0.1cm}
\caption{Diagrammatical representation of the tree-level contribution to
the partonic amplitude $\gamma \, \gamma^{\ast} \to q \, \bar q$ induced by two
electromagnetic currents. }
\label{fig:tree-level-pion-FF-LP}
\end{center}
\end{figure}

\subsection{QCD factorization of $F_{\gamma^{\ast} \gamma \to \pi^0} (Q^2)$ at tree level}

QCD factorization for the leading-twist contribution to the  $\gamma^{\ast} \gamma \to \pi^0$ form factor
at tree level can be established by inspecting the four-point QCD matrix element
\begin{eqnarray}
\Pi_{\mu}=\langle q(x\, p) \, \bar q (\bar x \, p) | j_{\mu}^{\rm em}  | \gamma (p^{\prime}) \rangle
\label{leading power correlator}
\end{eqnarray}
at leading order (LO) in $\alpha_s$, where  $x$ indicates the momentum fraction carried by
 the collinear quark of the pion and $\bar x \equiv 1-x$.
Computing the two diagrams displayed in  figure \ref{fig:tree-level-pion-FF-LP} yields
\begin{eqnarray}
\Pi_{\mu}^{(0)} &=& - \frac{i \, g_{\rm em}^2 \, (Q_u^2-Q_d^2)}{2 \, \sqrt{2} \, \, n \cdot p} \, \epsilon^{\nu}(p^{\prime}) \,
\left \{  \frac{\left [ \bar u(x \, p) \,  \gamma_{\mu, \perp}  \, \not \! \bar n \,\,
\gamma_{\nu, \perp}  \, \, v(\bar x \,  p)  \right ]}{\bar x}  \,
-  \frac{\left [  \bar u(x \, p) \,  \gamma_{\nu, \perp}  \, \not \! \bar n \,\,
\gamma_{\mu, \perp}  \, \, v(\bar x \,  p)  \right ] }{x}   \,   \right \}   \nonumber \\
&=& - \frac{i \, g_{\rm em}^2 \, (Q_u^2-Q_d^2)}{2 \, \sqrt{2} \, \, n \cdot p} \, \epsilon^{\nu}(p^{\prime}) \,
\left [  {1 \over \bar x^{\prime}}  \, \ast \, \langle O_{A, \, \mu \nu} (x, x^{\prime}) \rangle^{(0)}
- {1 \over x^{\prime}}   \, \ast \, \langle O_{B, \, \mu \nu} (x, x^{\prime}) \rangle^{(0)}   \right ] \,.
\end{eqnarray}
Here, $\langle O_{A, \, \mu \nu} \rangle^{(0)}$ and $\langle O_{B, \, \mu \nu} \rangle^{(0)}$
denote the tree-level partonic  matrix elements
of the  collinear operators $O_{A, \, \mu \nu}$ and $O_{B, \, \mu \nu}$ in soft-collinear effective theory (SCET)
\begin{eqnarray}
\langle O_{j, \, \mu \nu} (x, x^{\prime})  \rangle \equiv
\langle q(x\, p) \, \bar q (\bar x \, p) | O_{j, \, \mu \nu}(x^{\prime}) | 0 \rangle
=\bar \xi(x \, p)  \, \Gamma_{j, \, \mu \nu} \, \xi(\bar x \, p) \,\, \delta(x - x^{\prime}) + {\cal O}(\alpha_s) \,,
\end{eqnarray}
and  the convolution integration is represented by an asterisk.
The manifest definition of the SCET operator $ O_{j, \, \mu \nu}$ in the momentum  space are given by
\begin{eqnarray}
O_{j, \, \mu \nu}(x^{\prime})= {\bar n \cdot p \over 2 \pi}  \, \int d \tau  \, e^{i \, x^{\prime} \, \tau \, \bar n \cdot p} \,\,
\bar \xi (\tau \bar n) \, W_c(\tau \bar n,0) \, \Gamma_{j, \, \mu \nu}   \,  \xi(0) \,,
\end{eqnarray}
with the collinear Wilson line
\begin{eqnarray}
W_c(\tau \bar n,0) = {\rm P} \, \left \{ {\rm  Exp} \left [   i \, g_s \,
\int_{0}^{\tau} \, d \lambda \,  \bar n  \cdot A_{c}(\lambda \, \bar n) \right ]  \right \}
\end{eqnarray}
and
\begin{eqnarray}
\Gamma_{A, \, \mu \nu} = \gamma_{\mu, \perp}  \, \not \! \bar n \,\,
\gamma_{\nu, \perp}  \,, \qquad
\Gamma_{B, \, \mu \nu} =  \gamma_{\nu, \perp}  \, \not \! \bar n \,\,
\gamma_{\mu, \perp}   \,.
\end{eqnarray}

To facilitate the determination of the hard function entering the leading power factorization formula
of $F_{\gamma^{\ast} \gamma \to \pi^0} (Q^2)$, we employ the SCET operator basis
$\{ O_{1, \, \mu \nu}, \, O_{2, \, \mu \nu}, \, O_{E, \, \mu \nu} \}$  with
\begin{eqnarray}
\Gamma_{1, \, \mu \nu} = g_{\mu \nu}^{\perp}  \, \not \! \bar n  \,, \qquad
\Gamma_{2, \, \mu \nu} =  i \, \epsilon_{\mu \nu}^{\perp} \, \not \! \bar n  \, \gamma_5 \,,   \qquad
\Gamma_{E, \, \mu \nu} =  \not \! \bar n \left ( { [ \gamma_{\mu, \perp},  \gamma_{\nu, \perp}] \over 2 }
-  i \, \epsilon_{\mu \nu}^{\perp} \, \gamma_5 \right )\,,
\end{eqnarray}
where $O_{E, \, \mu \nu}$ is an evanescent operator vanishing in four dimensions and
\begin{eqnarray}
g^{\perp}_{\mu \nu} \equiv g_{\mu \nu}-\frac{n_{\mu} \bar n_{\nu}}{2} -\frac{n_{\nu} \bar n_{\mu}}{2} \,, \qquad
\epsilon_{\mu \nu}^{\perp}  \equiv {1 \over 2} \, \epsilon_{\mu \nu \alpha \beta} \bar n^{\alpha} \, n^{\beta}\,.
\end{eqnarray}
It is evident that the effective operator $O_{1, \, \mu \nu}$ cannot couple with a collinear pion state
due to the parity conservation. Taking advantages of the operator identities
\begin{eqnarray}
O_{A, \, \mu \nu} &=& - \left ( O_{1, \, \mu \nu} + O_{2, \, \mu \nu} + O_{E, \, \mu \nu} \right )  \,, \nonumber \\
O_{B, \, \mu \nu} &=& -  \left ( O_{1, \, \mu \nu} -  O_{2, \, \mu \nu} - O_{E, \, \mu \nu} \right )  \,,
\end{eqnarray}
we observe that the two tree-level diagrams in figure \ref{fig:tree-level-pion-FF-LP}
give rise to the identical contribution to the pion-photon
transition form factor and such observation can be further generalized to all orders in QCD applying the charge-conjugation transformation.

We now employ the operator matching equation with the evanescent  operator
\begin{eqnarray}
\Pi_{\mu} =\left [\frac{i \, g_{\rm em}^2 \, (Q_u^2-Q_d^2)}{2 \, \sqrt{2} \, \, n \cdot p} \,
 \epsilon^{\nu}(p^{\prime})  \right ]  \, \sum_i \, T_{i}(x^{\prime})   \, \ast \langle O_{i, \, \mu \nu}(x, x^{\prime}) \rangle \,,
 \label{matching equation}
\end{eqnarray}
and  expand all quantities to the tree level, yielding
\begin{eqnarray}
T^{(0)}_{1}(x^{\prime})= {1 \over x^{\prime}} - {1 \over \bar x^{\prime}}      \,,  \qquad
T^{(0)}_{2}(x^{\prime})=T^{(0)}_{E}(x^{\prime})={1 \over x^{\prime}} + {1 \over \bar x^{\prime}}  \,.
\end{eqnarray}
Utilizing the definition of the leading twist pion DA on the light cone
\begin{eqnarray}
\langle  \pi(p) |\bar \xi (y) \, W_c(y, 0) \, \gamma_{\mu} \, \gamma_5 \, \xi(0) |  0 \rangle
= -i \, f_{\pi} \, p_{\mu}  \, \int_0^1 \, d u \, e^{i\, u \, p \cdot y}  \, \phi_{\pi} (u, \mu)
+ {\cal O}(y^2) \,,
\end{eqnarray}
it is straightforward to derive the tree-level factorization formula of the $\pi^{0} \gamma^{\ast} \gamma $ form factor
\begin{eqnarray}
F_{\gamma^{\ast} \gamma \to \pi^0}^{\rm LP} (Q^2)= \frac{\sqrt{2} \, (Q_u^2-Q_d^2) \, f_{\pi}}{Q^2} \,
\int_0^1 \, d x \, T^{(0)}_{2}(x)\, \phi_{\pi} (x, \mu)
+ {\cal O}(\alpha_s)   \,.
\end{eqnarray}

\subsection{QCD factorization  of $F_{\gamma^{\ast} \gamma \to \pi^0} (Q^2)$  at ${\cal O} (\alpha_s)$}

\begin{figure}
\begin{center}
\includegraphics[width=1.0 \columnwidth]{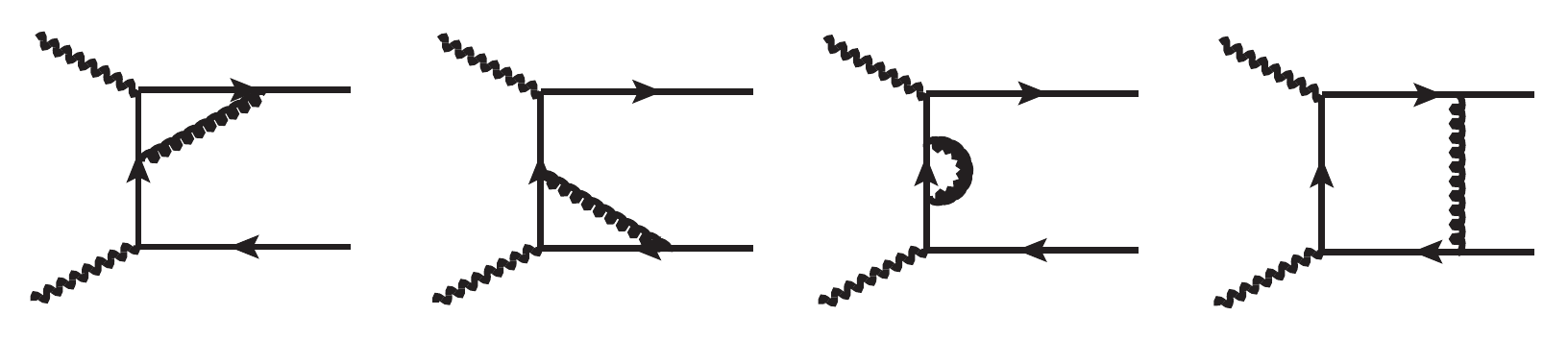} \\
(a) \hspace{3.5 cm} (b)  \hspace{3.5 cm} (c)  \hspace{3.5 cm} (d)
\vspace*{0.1cm}
\caption{Diagrammatical representation of the one-loop contribution to
the partonic amplitude $\gamma \, \gamma^{\ast} \to q \, \bar q$ induced by two
electromagnetic currents. The corresponding symmetric diagrams obtained by exchanging
the two photon states are not shown. }
\label{fig:one-loop-pion-FF-LP}
\end{center}
\end{figure}

We proceed to compute the NLO QCD correction to the four-point partonic amplitude $\Pi_{\mu}^{(1)}$
at leading power in $1/Q^2$ for the determination of the hard function $T_{2}$ at ${\cal O}(\alpha_s)$.
It needs to be stressed that the QCD matrix element $\Pi_{\mu}$ defined by two electromagnetic
currents is {\it independent} of the prescription of $\gamma_5$ in  the  $D$-dimensional space
and the renormalization scheme dependence of the perturbative matching coefficient $T_2$
comes solely from the radiative correction to the twist-2 pion DA $\phi_{\pi} (x, \mu)$, whose definition depends on
the precise treatment of the Dirac matrix $\gamma_5$ in dimensional regularization,
due to the infrared subtraction.
Evaluating the hard contribution to the one-loop diagrams displayed in figure \ref{fig:one-loop-pion-FF-LP}
with the method of regions \cite{Beneke:1997zp} immediately leads to
\begin{eqnarray}
\Pi_{\mu}^{(1a)} &=&  \frac{i \, g_{\rm em}^2 \, (Q_u^2-Q_d^2)}{2 \, \sqrt{2} \, \, n \cdot p}
{\alpha_s \, C_F \over 2 \pi} \, \, \epsilon^{\nu}(p^{\prime}) \,
\langle O_{2, \, \mu \nu}(x, x^{\prime}) \rangle^{(0)} \,\,     \nonumber \\
&&  \ast \, \bigg \{  {1 \over x^{\prime} \, \bar x^{\prime}} \, \left  [ - \left (\ln \bar x^{\prime}  + {x^{\prime} \over 2} \right ) \,
\left ( {1 \over \epsilon} + \ln {\mu^2 \over Q^2} \right ) \,
+ {1 \over 2} \, \ln \bar x^{\prime} \, \left ( \ln \bar x^{\prime} -2 - \bar x^{\prime}  \right )  \, - 2 \, x^{\prime}  \right ]
 + ...  \,,   \hspace{0.2 cm} \\
\Pi_{\mu}^{(1b)} &=&  \frac{i \, g_{\rm em}^2 \, (Q_u^2-Q_d^2)}{2 \, \sqrt{2} \, \, n \cdot p} \,
{\alpha_s \, C_F \over 2 \pi} \, \epsilon^{\nu}(p^{\prime}) \,
\langle O_{2, \, \mu \nu}(x, x^{\prime}) \rangle^{(0)} \,\,   \nonumber \\
&&  \ast \, \left \{  {1 \over \bar x^{\prime}} \, \left  [ - {1 \over 2} \,
\left ( {1 \over \epsilon} + \ln {\mu^2 \over Q^2} - \ln \bar x^{\prime} \right ) \, - 2  \right ]
 \right \}  + ...  \,,  \\
\Pi_{\mu}^{(1c)} &=&  - \frac{i \, g_{\rm em}^2 \, (Q_u^2-Q_d^2)}{2 \, \sqrt{2} \, \, n \cdot p}
\, {\alpha_s \, C_F \over 4 \pi}  \, \epsilon^{\nu}(p^{\prime}) \,
\langle O_{2, \, \mu \nu}(x, x^{\prime}) \rangle^{(0)} \,\,
\ast \, \left \{  {1 \over \bar x^{\prime}} \, \left  [ {1 \over \epsilon} + \ln {\mu^2 \over \bar x^{\prime} \,  Q^2}  \, + 1  \right ]
\right \}  + ...  \,,  \hspace{0.6 cm} \\
\Pi_{\mu}^{(1d)} &=&  \frac{i \, g_{\rm em}^2 \, (Q_u^2-Q_d^2)}{2 \, \sqrt{2} \, \, n \cdot p} \,
{\alpha_s \, C_F \over 2 \pi} \, \epsilon^{\nu}(p^{\prime}) \,
\langle O_{2, \, \mu \nu}(x, x^{\prime}) \rangle^{(0)} \,\,    \nonumber \\
&&   \ast \, \left \{  {\ln \bar x^{\prime}  \over x^{\prime}} \, \left  [ {1 \over \epsilon} + \ln {\mu^2 \over Q^2}  \,
-{1 \over 2} \, \ln \bar x^{\prime}  + 5  \right ] \right \}  + ...  \,,
\end{eqnarray}
where the ellipses represent terms proportional to $\langle O_{1, \, \mu \nu}(x, x^{\prime}) \rangle^{(0)} $
and $\langle O_{E, \, \mu \nu}(x, x^{\prime}) \rangle^{(0)}$.
Adding up different pieces together we can readily obtain the  QCD matrix element $\Pi_{\mu}$ at ${\cal O}(\alpha_s)$
\begin{eqnarray}
\Pi_{\mu}^{(1)}= \frac{i \, g_{\rm em}^2 \, (Q_u^2-Q_d^2)}{2 \, \sqrt{2} \, \, n \cdot p} \, \epsilon^{\nu}(p^{\prime}) \,
\langle O_{2, \, \mu \nu}(x, x^{\prime}) \rangle^{(0)}  \, \ast \,  A_{2, \rm hard}^{(1)}(x^{\prime})  + ... \,,
\end{eqnarray}
where the $\gamma_5$-prescription independent amplitude $A_{2, \rm hard}^{(1)}$ reads
\begin{eqnarray}
A_{2, \rm hard}^{(1)}(x^{\prime}) &=& {\alpha_s \, C_F \over 4 \pi} \,
\bigg \{ {1 \over \bar x^{\prime}}  \, \left [ - \left (2 \, \ln \bar x^{\prime} + 3 \right ) \,
\left ( {1 \over \epsilon} + \ln {\mu^2 \over Q^2}   \right )  + \ln^2 \bar x^{\prime}
+ 7 \, { \bar  x^{\prime} \, \ln  \bar x^{\prime}  \over  x^{\prime} }  - 9  \right ]  \nonumber \\
&& \hspace{1.5 cm}  + \left (  x^{\prime} \leftrightarrow  \bar x^{\prime}  \right )  \bigg \}  \,.
\label{one-loop hard QCD amplitude}
\end{eqnarray}
Expanding the matching equation (\ref{matching equation}) to the one-loop order yields
\begin{eqnarray}
&& \left [\frac{i \, g_{\rm em}^2 \, (Q_u^2-Q_d^2)}{2 \, \sqrt{2} \, \, n \cdot p} \,
\epsilon^{\nu}(p^{\prime})  \right ]  \, \sum_i \, A_{i}^{(1)}(x^{\prime})   \, \ast
\langle O_{i, \, \mu \nu}(x, x^{\prime}) \rangle^{(0)} \nonumber \\
&& =\left [\frac{i \, g_{\rm em}^2 \, (Q_u^2-Q_d^2)}{2 \, \sqrt{2} \, \, n \cdot p} \,
\epsilon^{\nu}(p^{\prime})  \right ]  \,   \sum_i \,
\left [ T_i^{(1)}(x^{\prime}) \ast   \langle O_{i, \, \mu \nu}(x, x^{\prime}) \rangle^{(0)}
+ T_i^{(0)}(x^{\prime}) \ast   \langle O_{i, \, \mu \nu}(x, x^{\prime}) \rangle^{(1)}\right ]  \,. \hspace{1.0 cm}
\label{one-loop matching condition}
\end{eqnarray}
Now we are in a position to derive the master formula for the one-loop perturbative matching coefficient $T_i^{(1)}$
by implementing both the ultraviolet (UV) renormalization and the infrared (IR) subtraction.
Following the strategy presented in \cite{Beneke:2005vv} the UV renormalized matrix element
of the SCET operator $O_{i, \, \mu \nu}$ at ${\cal O}(\alpha_s)$ is given by
\begin{eqnarray}
\langle O_{i, \, \mu \nu} \rangle^{(1)} = \sum_j \,
\left [ M_{i j, {\rm bare}}^{(1) R}  + Z_{i j}^{(1)} \right ] \, \ast \, \langle O_{j, \, \mu \nu} \rangle^{(0)} \,,
\label{one-loop SCET operator matrix element}
\end{eqnarray}
where the bare matrix element $M_{i j, {\rm bare}}^{(1)}$ depends on the IR regularization scheme $R$.
Applying the dimensional regularization for both the UV and IR divergences,
the bare matrix element $M_{i j, {\rm bare}}^{(1)} $ vanishes due to  scaleless integrals entering the relevant one-loop computation.
Inserting (\ref{one-loop SCET operator matrix element}) into  (\ref{one-loop matching condition}) and comparing the coefficient
of $\langle O_{2, \, \mu \nu} \rangle^{(0)}$ give rise to
\begin{eqnarray}
T_2^{(1)} = A_2^{(1)} - \sum_i \, T_i^{(0)}  \ast Z_{i 2}^{(1)}  \,.
\label{infrared subtraction: preliminary}
\end{eqnarray}
It is evident that  the  SCET operators $O_{1, \, \mu \nu}$ and $O_{2, \, \mu \nu}$ cannot mix into each other
under QCD renormalization due to the parity conservation,  hence $Z_{1 2}^{(1)}=0$.
In addition,  the IR subtraction term $T_2^{(0)}  \ast Z_{2 2}^{(1)} $ will remove the collinear contribution
to the QCD amplitude $\Pi_{\mu}$ at one loop so that the matching coefficient $T_2^{(1)}$ only encodes the information
of strong interaction dynamics at the hard scale.
Technically, the collinear subtraction has been automatically implemented in the above computation of the QCD matrix
element $\Pi_{\mu}$, since only the hard contribution computed with  the expansion by regions enters the expression
of $A_{2, \rm hard}^{(1)}$ displayed in (\ref{one-loop hard QCD amplitude}).

We are now ready to discuss the renormalization constant $Z_{E2}$ of the evanescent operator $O_{E, \, \mu \nu}$
for the derivation of the final result of the matching coefficient $T_2^{(1)}$.
Applying the renormalization prescription that the IR finite matrix element of the evanescent operator
$\langle O_{E, \, \mu \nu} \rangle $ vanishes with  dimensional regularization applied only to
the UV divergences and with the IR singularities regularized by any parameter other than
the dimensions of spacetime \cite{Dugan:1990df,Herrlich:1994kh} and
making use of the identity (\ref{one-loop SCET operator matrix element}) yield
\begin{eqnarray}
Z_{E2}^{(1)}=-M_{E2}^{(1) \rm off} \,.
\label{renormalization constant of evanescent operator}
\end{eqnarray}
The one-loop matching coefficient of the physical operator $O_{2, \, \mu \nu}$ can be readily obtained
by substituting  (\ref{renormalization constant of evanescent operator}) into (\ref{infrared subtraction: preliminary})
\begin{eqnarray}
T_2^{(1)} = A_2^{(1)} -  T_2^{(0)} \ast Z_{2 2}^{(1)}  + T_E^{(0)} \ast  M_{E2}^{(1) \rm off}
=  A_{2, \rm hard}^{(1)}  +   T_E^{(0)} \ast  M_{E2}^{(1) \rm off}  \,.
\label{infrared subtraction: final}
\end{eqnarray}

\begin{figure}
\begin{center}
\includegraphics[width=0.90 \columnwidth]{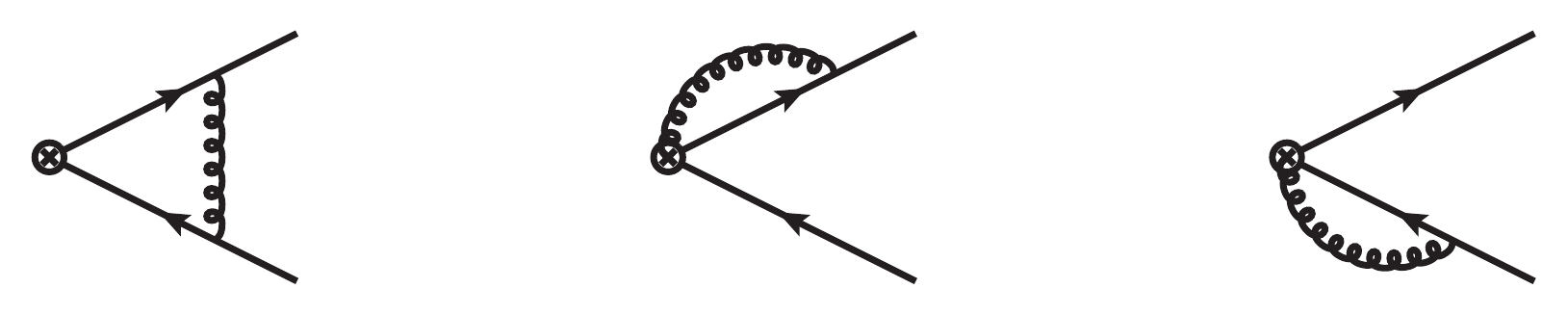} \\
(a) \hspace{5 cm} (b)  \hspace{5 cm} (c)
\vspace*{0.1cm}
\caption{The one-loop SCET diagrams contributing to $M_{E2}^{(1) \rm off}$
and the vertex ``$\otimes$" indicating an insertion of the evanescent operator $O_{E, \, \mu \nu}$. }
\label{fig:one-loop-OE-LP}
\end{center}
\end{figure}

The one-loop contribution to the matrix element of the evanescent operator $O_{E, \, \mu \nu}$ depends on the
renormalization prescription of $\gamma_5$ in the $D$-dimensional space.
We will employ both the NDR and HV schemes of $\gamma_5$ below for the illustration of the prescription independence
of the factorization formula of $F_{\gamma^{\ast} \gamma \to \pi^0} (Q^2)$  at ${\cal O} (\alpha_s)$ and
at leading power in $1/Q^2$.  This amounts to showing that the renormalization scheme dependence of
the short-distance coefficient function cancels against that of the twist-2 pion DA precisely.
Evaluating the SCET diagrams displayed in figure \ref{fig:one-loop-OE-LP} with the Wilson-line Feynman rules,  we find that
only a single diagram \ref{fig:one-loop-OE-LP}(a) can generate a nonvanishing contribution to
$M_{E2}^{(1) \rm off}$ using the NDR scheme of $\gamma_5$,
which turns out to be proportional to the spin-dependent term of the Brodsky-Lepage evolution kernel
\cite{Lepage:1980fj,Efremov:1979qk}.  The manifest expression of the infrared subtraction term
$T_E^{(0)} \ast  M_{E2}^{(1) \rm off}$  is then given by
\begin{eqnarray}
T_E^{(0)} \ast  M_{E2}^{(1) \rm off} \, \big |_{\rm NDR} &=&  {\alpha_s \, C_F \over 2 \pi} \,
\int_0^1 \, d y \, \left ( {1 \over y} + {1 \over \bar y} \right ) \, 4 \,
\left [ {\bar y \over \bar x^{\prime}} \, \theta(y-x^{\prime})
+  {y \over x^{\prime}} \, \theta(x^{\prime} - y)  \right ] \nonumber \\
&=&   {\alpha_s \, C_F \over 2 \pi} \, (-4) \,
\left ( {\ln \bar x^{\prime}  \over  x^{\prime}}  + {\ln x^{\prime}  \over  \bar x^{\prime}}   \right )  \,.
\label{IR subtraction: NDR}
\end{eqnarray}
However, computing the one-loop matrix element of the evanescent operator $O_{E, \, \mu \nu}$ with the HV scheme
of $\gamma_5$ leads to
\begin{eqnarray}
T_E^{(0)} \ast  M_{E2}^{(1) \rm off} \, \big |_{\rm HV} &=&  0 \,.
\label{IR subtraction: HV}
\end{eqnarray}
Inserting (\ref{one-loop hard QCD amplitude}), (\ref{IR subtraction: NDR}) and  (\ref{IR subtraction: HV})
into the master formula (\ref{infrared subtraction: final}), we obtain
\begin{eqnarray}
T_2^{(1)}(x^{\prime}, \mu) &=& {\alpha_s \, C_F \over 4 \pi} \,
\bigg \{ {1 \over \bar x^{\prime}}  \, \left [ - \left (2 \, \ln \bar x^{\prime} + 3 \right ) \,
\ln {\mu^2 \over Q^2} + \ln^2 \bar x^{\prime}
+ \delta \, { \bar  x^{\prime} \, \ln  \bar x^{\prime}  \over  x^{\prime} }  - 9  \right ]
+ \left (  x^{\prime} \leftrightarrow  \bar x^{\prime}  \right )  \bigg \}  \,, \hspace{0.5 cm}
\label{one-loop hard function}
\end{eqnarray}
where the renormalization scheme dependent parameter $\delta$ is given by
\begin{eqnarray}
\delta = \left\{
\begin{array}{l}
-1 \,,  \qquad
[{\rm NDR \,\,scheme}] \vspace{0.4 cm} \\
+ 7 \,.  \qquad
[{\rm HV \,\,scheme}]
\end{array}
 \hspace{0.5 cm} \right.
\end{eqnarray}
Our result of $T_2^{(1)}$ in the NDR scheme is identical to that  presented in \cite{Braaten:1982yp} using the trace formalism.

We  now aim at demonstrating the renormalization prescription independence of the one-loop factorization
formula for the pion-photon form factor
\begin{eqnarray}
F_{\gamma^{\ast} \gamma \to \pi^0}^{\rm LP} (Q^2)= \frac{\sqrt{2} \, (Q_u^2-Q_d^2) \, f_{\pi}}{Q^2} \,
\int_0^1 \, d x \, \left[  T^{(0)}_{2}(x)  + T^{(1), \, \Delta}_{2}(x, \mu) \right ] \, \phi_{\pi}^{\Delta} (x, \mu)
+ {\cal O}(\alpha_s^2)   \,,
\label{one-loop factorization formula at LP}
\end{eqnarray}
where the superscript ``$\Delta$" indicates the $\gamma_5$-scheme in dimensional regularization.
Taking advantage of the relation of the twist-2 pion DA between the NDR and HV schemes
\begin{eqnarray}
\phi_{\pi}^{\rm HV} (x, \mu) = \int_0^1 d y \,  Z_{\rm HV}^{-1}(x, y, \mu) \, \phi_{\pi}^{\rm NDR} (y, \mu)  \,,
\end{eqnarray}
where the finite renormalization kernel $Z_{\rm HV}^{-1}$ is given by \cite{Melic:2001wb}
\begin{eqnarray}
Z_{\rm HV}^{-1}(x, y, \mu) =  \delta(x-y) + {\alpha_s \, C_F \over 2 \pi} \, 4 \,
\left [ {x \over y} \, \theta(y-x) + {\bar x \over \bar y} \, \theta(x-y) \right ]  + {\cal O}(\alpha_s^2)  \,.
\end{eqnarray}
It is then straightforward to show that
\begin{eqnarray}
&& \int_0^1 d x \, T^{(0)}_{2}(x) \, \left [  \phi_{\pi}^{\rm HV} (x, \mu)  - \phi_{\pi}^{\rm NDR} (x, \mu)  \right ] \nonumber \\
&& =  {\alpha_s \, C_F \over 2 \pi} \, (-4) \,  \int_0^1 \, d y \,
\left ( {\ln \bar y  \over  y}  + {\ln y  \over  \bar y}   \right )  \, \phi_{\pi}^{\rm NDR} (x, \mu) + {\cal O}(\alpha_s^2)  \,,
\end{eqnarray}
which cancels against the renormalization scheme dependence of the NLO hard function $T^{(1), \, \Delta}_{2}(x, \mu)$ as displayed in (\ref{one-loop hard function}). We emphasize again that the $\gamma_5$-prescription independence of the leading power factorization formula
for $F_{\gamma^{\ast} \gamma \to \pi^0} (Q^2)$ stems from the fact that the QCD matrix element $\langle q(x\, p) \, \bar q (\bar x \, p) | j_{\mu}^{\rm em}  | \gamma (p^{\prime}) \rangle$ itself is free of the $\gamma_5$ ambiguity in dimensional regularization
and both the NDR and HV prescriptions can be employed to construct QCD factorization theorems for hard   processes
provided that the corresponding matching coefficients are computed in an appropriate way without overlooking the potential evanescent operators.

The renormalization scale independence of the factorization formula (\ref{one-loop factorization formula at LP})
 can be readily verified by making use of the evolution equation of  the pion DA $\phi_{\pi} (x, \mu)$
\begin{eqnarray}
\mu^2 \, {d \over d \mu^2} \, \phi_{\pi} (x, \mu) = \int_0^1 \, d y \, V(x, y) \, \phi_{\pi} (y, \mu) \,,
\label{RGE of pion DA}
\end{eqnarray}
where the evolution kernel $ V(x, y)$ can be expanded perturbatively in QCD
\begin{eqnarray}
V(x, y) = \sum_{n=0}  \, \left ( {\alpha_s \over 4 \pi} \right )^{n+1} \, \left [ V_n(x,y) \right ]_{+}\,,
\end{eqnarray}
with the ``plus" function defined as
\begin{eqnarray}
\left [ f(x,y) \right ]_{+} = f(x,y) - \delta(x-y) \, \int_0^1 \, d t \, f(t, y) \,,
\end{eqnarray}
and the LO Brodsky-Lepage kernel given by \cite{Lepage:1980fj,Efremov:1979qk}
\begin{eqnarray}
V_0(x, y) = 2\, C_F \, \left [ {1 - x \over 1 -y} \,
\left ( 1 + {1 \over x - y} \right ) \, \theta(x-y)
+ {x \over y} \, \left ( 1 + {1 \over y - x} \right ) \, \theta(y-x)   \right ]  \,.
\end{eqnarray}
It is appropriate to point out that the one-loop evolution kernel $V_0(x, y)$ is independent
of the $\gamma_5$ prescription in  the complex $D$-dimensional space, however, the two-loop
evolution kernel $V_1(x, y)$ does depend on the renormalization scheme.
Applying the renormalization-group (RG) equation (\ref{RGE of pion DA}) then leads to
\begin{eqnarray}
{d \over d \ln \mu}  \, F_{\gamma^{\ast} \gamma \to \pi^0} (Q^2)= {\cal O}(\alpha_s^2) \,.
\end{eqnarray}

We further turn to sum the parametrically large logarithms of $Q^2/\mu^2$ in the short-distance
function at next-to-leading-logarithmic (NLL) accuracy employing the standard RG approach in
the momentum space.  Technically, the desired NLL resummation can be readily achieved
by setting the factorization scale of order $\mu \sim \sqrt{Q^2}$ and evolving the leading twist
pion DA up to that scale at two loops. The NLO evolution kernel $V_1(x, y)$ was first
obtained with the diagrammatic approach in the light-cone gauge \cite{Sarmadi:1982yg,Dittes:1983dy},
then in the Feynman gauge \cite{Katz:1984gf,Mikhailov:1984ii},
and was further reconstructed \cite{Belitsky:1999gu} based upon the knowledge of the conformal anomalies
and the available forward DGLAP splitting functions at ${\cal O}(\alpha_s^2)$.
The two-loop evolution potential $V_1(x, y)$  can be organized as
\begin{eqnarray}
V_1(x, y) = 2\,N_f \, C_F \, V_N(x, y) + 2 \, C_F \, C_A \, V_G(x, y) + C_F^2 \, V_F(x, y) \,,
\end{eqnarray}
where $N_f$ is the number of the active quark flavours. The explicit expressions of the kernels
$V_N$, $V_G$ and $V_F$ are given by  \cite{Mikhailov:1985cm}
\begin{eqnarray}
V_N(x, y)  &=&   - {2 \over 3} \, \theta(y-x) \, \left [ {5 \over 3} \, F(x, y) +  {x \over y}+
 \, F(x, y) \, \ln {x \over y} \right ]
+ \left (x \leftrightarrow \bar x \,, y \leftrightarrow \bar y \right ) \,,  \\
V_G(x, y)  &=& \left \{ \theta(y-x) \,   \left [ {67 \over 9} \,  F(x, y) + {17 \over 3}  \, {x \over y}
+ {11 \over 3} \, F(x, y) \, \ln {x \over y} \right ] + H(x, y) \right \} \nonumber \\
&& + \left (x \leftrightarrow \bar x \,, y \leftrightarrow \bar y \right ) \,, \\
V_F(x, y)  &=&   4 \, \bigg \{ \theta(y-x) \,  \bigg [ -{\pi^2 \over 3} \,  F(x, y) + {x \over y}
- \left ( {3 \over 2} \,  F(x, y)  - {x \over 2 \, \bar y} \right )   \, \ln {x \over y}  \nonumber \\
&& - \left ( F(x, y) - F(\bar x, \bar y) \right ) \, \ln {x \over y} \, \ln \left (1-{x \over y} \right )
+ \left ( F(x, y) + {x \over 2 \, \bar y} \right ) \,  \ln^2 {x \over y}   \bigg ] \nonumber \\
&& - {x \over 2 \, \bar y} \, \ln x \, \left ( 1 + \ln x - 2 \, \ln \bar x  \right )
- H(x, y)  \bigg \} + \left (x \leftrightarrow \bar x \,, y \leftrightarrow \bar y \right ) \,,
\end{eqnarray}
where we have introduced the functions $F(x, y)$ and $H(x, y)$ as follows
\begin{eqnarray}
F(x, y) &=& {x \over y} \, \left ( 1 + {1 \over y - x} \right ) \,,  \\
H(x, y) &=&  \theta(x -\bar y)  \, \bigg[ 2 \, \left ( F(x, y) -  F(\bar x, \bar y) \right ) \,
{\rm Li}_2 \left ({1 - {x \over y}} \right ) - 2 \,  F(x, y) \, \ln x \, \ln y   \nonumber \\
&& + \left ( F(x, y) -  F(\bar x, \bar y) \right ) \, \ln^2 y  \bigg ]
+ 2 \, F(x, y)  \,{\rm Li}_2 (\bar y) \, \left [ \theta(x -\bar y)  - \theta(y-x)   \right ] \nonumber \\
&& + 2 \, \theta(y-x) \,  F(\bar x, \bar y) \,  \ln y  \, \ln {\bar x}
- 2 \, F(x, y)  \,{\rm Li}_2 (x) \,  \left [ \theta(x -\bar y)   - \theta(x - y)  \right ]   \,.
\end{eqnarray}
In order to perform the NLL QCD resummation, it turns out to be convenient to adopt the Gegenbauer
expansion of the pion DA
\begin{eqnarray}
\phi_{\pi}(x ,\mu) = 6 \, x \, \bar x \, \sum_{n=0}^{\infty} \, a_n(\mu) \, C_n^{3/2}(2 x-1)\,,
\label{Gegenbauer expansion of pion DA}
\end{eqnarray}
where the LO coefficient $a_0(\mu)=1$ is renormalization invariant due to the normalization condition.
The exact solution to the two-loop RG equation (\ref{RGE of pion DA}) can be constructed from the forward
anomalous dimensions and the special conformal anomaly matrix  in the Gegenbauer moment space
\cite{Mueller:1993hg,Mueller:1994cn}, and we obtain (see also \cite{Agaev:2010aq})
\begin{eqnarray}
a_n(\mu) =  E_{V, n}^{\rm NLO}(\mu, \mu_0) \, a_n(\mu_0)
+  {\alpha_s(\mu) \over 4 \pi} \, \sum_{k=0}^{n-2} \,  E_{V, n}^{\rm LO}(\mu, \mu_0) \,
d_{V, n}^{k}(\mu, \mu_0) \, a_k(\mu_0)  \,,
\label{NLL evolution of pion moments}
\end{eqnarray}
where both $n$ and $k$ are non-negative even integers and the explicit expressions of
$E_{V, n}^{\rm NLO}$ and $d_{V, n}^{k}$ are collected in Appendix \ref{app:two-loop evolutions}.
Inserting (\ref{Gegenbauer expansion of pion DA}) into (\ref{one-loop factorization formula at LP})
and employing the  technique developed in \cite{Grossmann:2015lea},  we obtain
\begin{eqnarray}
F_{\gamma^{\ast} \gamma \to \pi^0}^{\rm LP} (Q^2)=  \frac{3 \, \sqrt{2} \, (Q_u^2-Q_d^2)}{Q^2} \,
f_{\pi} \, \sum_{n=0}^{\infty} \, a_{n}(\mu) \, C_n(Q^2, \mu)  + {\cal O}(\alpha_s^2) \,,
\label{NLL resummation for the LP effect}
\end{eqnarray}
where the hard coefficient $C_n(Q^2, \mu)$ in the NDR scheme of $\gamma_5$ is given by
\begin{eqnarray}
C_n(Q^2, \mu) &=&  1 + {\alpha_s(\mu) \, C_F \over 4\, \pi} \,
\bigg \{  \left [ 4 \, H_{n+1}  - {3\, n \, (n+3) + 8 \over (n+1)(n+2)} \right ] \, \ln {\mu^2 \over Q^2}
+ \, 4 \, H_{n+1}^2  - 4 \, { H_{n+1} + 1 \over (n+1)(n+2)}  \nonumber \\
&& +  \, 2 \, \left [ {1 \over (n+1)^2} + {1 \over (n+2)^2} \right ]
+ \, 3 \,  \left [ {1 \over (n+1)} - {1 \over (n+2)} \right ]   - 9 \bigg \} \,,
\end{eqnarray}
with the harmonic number defined as $ H_{n} = \Sigma_{k=1}^{k=n} \,\, 1/k $.

\section{The subleading-power correction from photon DAs}
\label{sect:subleading-power effect}

In this section we aim at evaluating the power suppressed contribution to the pion-photon
form factor due to the hadronic photon effect at ${\cal O}(\alpha_s)$ with the LCSR approach.
To this end, we construct the following vacuum-to-photon correlation function
\begin{eqnarray}
G_{\mu}(p^{\prime}, q) &=& \int d^4 z \, e^{-i \, q \cdot z}  \,
\langle 0 | {\rm T} \left \{ j_{\mu, \perp}^{\rm em}(z), j_{\pi}(0) \right \} | \gamma(p^{\prime}) \rangle  \,
\nonumber \\
&=& - g_{\rm em}^2 \, \epsilon_{\mu \nu \alpha \beta}^{\perp} \, q^{\alpha} \, p^{\prime \beta} \,
\epsilon_{\nu}(p^{\prime}) \, G(p^2, Q^2) \,,
\label{vacuum to photon correlator}
\end{eqnarray}
defined with an electromagnetic current (\ref{em current definition}) carrying a four-momentum $q_{\mu}$
and a pion interpolating current $j_{\pi}$ whose explicit structure is as follows
\begin{eqnarray}
j_{\pi} = {1 \over \sqrt{2}} \, \left ( \bar u \, \gamma_5 \, u
-  \bar d \, \gamma_5 \, d  \right )   \,.
\end{eqnarray}
Here we have introduced the convention
$\epsilon_{\mu \nu \alpha \beta}^{\perp} \equiv g_{\mu}^{\rho \, \perp}  \, \epsilon_{\rho \nu \alpha \beta}$.
Following the standard strategy, the primary task for the sum-rule construction consists in the demonstration of
QCD factorization for the considered correlation function (\ref{vacuum to photon correlator}) at space-like interpolating momentum
$p=p^{\prime} + q$. In contrast to the factorization proof of the leading power contribution presented in Section
\ref{sect:leading-power effect}, the QCD matrix element (\ref{vacuum to photon correlator}) itself
depends on the $\gamma_5$ prescription in $D$-dimensional space manifestly. 
We will employ both the NDR and HV schemes of the Dirac matrix $\gamma_5$ to establish the QCD factorization
formula of the transition amplitude  (\ref{vacuum to photon correlator}) at ${\cal O}(\alpha_s)$ and then derive the
NLL resummation improved LCSR for the hadronic photon correction to the $\pi^0 \gamma^{\ast} \gamma$ form factor.
Furthermore, the power counting rule for the external momenta
\begin{eqnarray}
| n \cdot p | \sim \bar n \cdot p \sim n \cdot p^{\prime} \sim {\cal O}(\sqrt{Q^2}) \,,
\end{eqnarray}
will  be adopted to determine the perturbative matching coefficient entering the factorization formula
of $G_{\mu}(p^{\prime}, q) $  to the one-loop order.

\subsection{The hadronic photon effect at tree level}

\begin{figure}
\begin{center}
\includegraphics[width=0.8 \columnwidth]{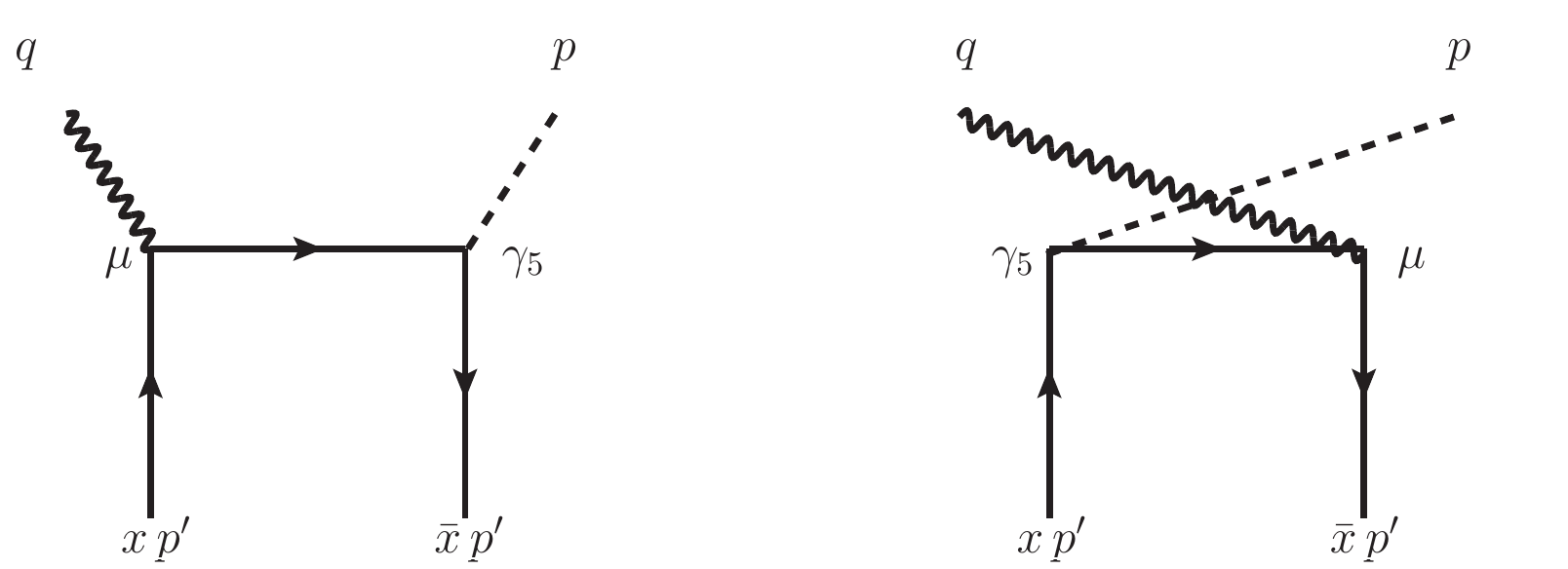} \\
(a) \hspace{7 cm} (b)
\vspace*{0.1cm}
\caption{Diagrammatical representation of the tree-level contribution to
the QCD amplitude $\widetilde{\Pi}_{\mu}$ defined in (\ref{vacuum to photon correlator: partonic}). }
\label{fig:tree-level-pion-FF-NLP}
\end{center}
\end{figure}

QCD factorization for the correlation function (\ref{vacuum to photon correlator}) at tree level can be established
by investigating the following four-point QCD amplitude
\begin{eqnarray}
\widetilde{\Pi}_{\mu} &=& \int d^4 z \, e^{-i \, q \cdot z}  \,
\langle 0 |{\rm T} \left \{ j_{\mu, \perp}^{\rm em}(z), j_{\pi}(0) \right  \} | q(x\, p^{\prime})
\, \bar q (\bar x \, p^{\prime}) \rangle \,
\label{vacuum to photon correlator: partonic}
\end{eqnarray}
at LO in $\alpha_s$.
Evaluating the two diagrams in figure \ref{fig:tree-level-pion-FF-NLP} leads to
\begin{eqnarray}
\widetilde{\Pi}_{\mu}^{(0)} &=& - {i \, g_{\rm em} \over 2 \, \sqrt{2}} \, {\bar n \cdot p \over Q^2} \,
\left [ {1 \over x \, r + \bar x}  + {1 \over \bar x \, r + x}   \right ]\, \sum_{q=u\,, d}
\eta_q \, Q_q \, \bar q(\bar x \,p^{\prime} ) \, \gamma_5 \, \not \! n  \,\, \gamma_{\mu, \perp}\, q(x \,p^{\prime} ) \nonumber \\
&=& - {i \, g_{\rm em} \over 2 \, \sqrt{2}} \, {\bar n \cdot p \over Q^2} \,  \sum_{q=u\,, d} \eta_q \, Q_q \,
\left [ {1 \over x^{\prime} \, r + \bar x^{\prime}}  + {1 \over \bar x^{\prime} \, r + x^{\prime}}   \right ]
\, \ast \, \langle  \widetilde{O}_{A, \mu}(x, x^{\prime})  \rangle^{(0)}  \,,
\label{NLP correlator at tree level}
\end{eqnarray}
where $r=-p^2/Q^2$, $\eta_u=1$ and $\eta_d=-1$.
The partonic matrix element of the (anti)-collinear SCET operator $\widetilde{O}_{A, \mu}$ at tree level is given by
\begin{eqnarray}
\langle  \widetilde{O}_{j, \mu}(x, x^{\prime})  \rangle \equiv
\langle 0 |\widetilde{O}_{j, \mu}(x^{\prime})| q(x\, p^{\prime})
\, \bar q (\bar x \, p^{\prime}) \rangle
= \bar \chi(\bar x \,p^{\prime})   \, \widetilde{\Gamma}_{j, \, \mu} \, \chi(x \,p^{\prime})\,
\delta(x- x^{\prime}) +  {\cal O}(\alpha_s)  \,.
\end{eqnarray}
The explicit definition of the (anti)-collinear operator $\widetilde{O}_{j, \mu}$
in the momentum space is
\begin{eqnarray}
\widetilde{O}_{j, \mu}(x^{\prime}) =  {n \cdot p^{\prime} \over 2 \pi}  \,
\int d \tau  \, e^{i \, x^{\prime} \, \tau \, n \cdot p^{\prime}} \,\,
\bar \chi (0) \, W_{\bar c}(0, \tau n) \, \widetilde{\Gamma}_{j, \, \mu}   \, \chi(\tau  n) \,,
\end{eqnarray}
where we have suppressed the flavour indices of $\widetilde{O}_{j, \mu}$ for brevity,
 $\widetilde{\Gamma}_{A, \, \mu} = \gamma_5  \not \! n  \,\, \gamma_{\mu, \perp}$ and
the corresponding  Wilson line is defined as
\begin{eqnarray}
W_{\bar c}(0, \tau n) &=& {\rm P} \, \left \{ {\rm  Exp} \left [ -  i \, g_s \,
\int_{0}^{\tau} \, d \lambda \,   n  \cdot A_{\bar c}(\lambda \, n) \right ]  \right \} \,.
\end{eqnarray}

To achieve the hard-collinear factorization for the correlation function (\ref{vacuum to photon correlator}),
we introduce  the SCET operator basis $\{ \widetilde{O}_{1, \mu},  \widetilde{O}_{E, \mu}\}$ with
\begin{eqnarray}
\widetilde{\Gamma}_{1, \, \mu} = {n^{\alpha}   \over 2} \, \epsilon_{\mu \nu \alpha \beta}^{\perp}  \,
\, \sigma^{\nu \beta} \,, \qquad
\widetilde{\Gamma}_{E, \, \mu} = \gamma_5 \, \not \! n  \,\, \gamma_{\mu, \perp} -
{n^{\alpha}   \over 2} \, \epsilon_{\mu \nu \alpha \beta}^{\perp}  \,
\, \sigma^{\nu \beta}  \,,
\end{eqnarray}
where $\widetilde{O}_{E, \mu}$ is evidently an evanescent operator.
Applying the operator matching equation including the evanescent operator
\begin{eqnarray}
\widetilde{\Pi}_{\mu} = - {i \, g_{\rm em} \over 2 \, \sqrt{2}} \, {\bar n \cdot p \over Q^2} \,  \sum_{q=u\,, d} \eta_q \, Q_q \,
\sum_i \, \widetilde{T}_{i}(x^{\prime})
\, \ast \, \langle  \widetilde{O}_{i, \mu}(x, x^{\prime})  \rangle \,,
\end{eqnarray}
and expanding all  quantities to the tree level, we can readily find that
\begin{eqnarray}
\widetilde{T}^{(0)}_{1}(x^{\prime}) = \widetilde{T}^{(0)}_{E}(x^{\prime})
= {1 \over x^{\prime} \, r + \bar x^{\prime}}  + {1 \over \bar x^{\prime} \, r + x^{\prime}}  \,.
\label{tree level hard function}
\end{eqnarray}
Making use of the leading twist DA of the photon defined in \cite{Ball:2002ps}
\begin{eqnarray}
&& \langle 0 |\bar \chi (0) \, W_{\bar c}(0, y) \, \sigma_{\alpha \beta} \, \chi (y) | \gamma(p^{\prime})  \rangle \nonumber \\
&& = i\, g_{\rm em} \, Q_q  \, \chi(\mu) \, \langle \bar q q \rangle (\mu)  \,
\left [   p^{\prime}_{\beta} \, \epsilon_{\alpha}(p^{\prime}) -
p^{\prime}_{\alpha} \, \epsilon_{\beta}(p^{\prime})\right ]   \,
\int_0^1 \, d u \, e^{- i u \, p^{\prime} \cdot y}   \, \phi_{\gamma}(u, \mu) \,,
\end{eqnarray}
the tree-level factorization formula of the form factor $G(p^2, Q^2)$ can be written as
\begin{eqnarray}
G(p^2, Q^2) = - {Q_u^2 - Q_d^2 \over \sqrt{2} \, Q^2} \,
\chi(\mu) \, \langle \bar q q \rangle (\mu) \,
\int_0^1 dx \, \widetilde{T}^{(0)}_{1}(x)  \,  \phi_{\gamma}(x, \mu) + {\cal O}(\alpha_s) \,,
\end{eqnarray}
where the magnetic susceptibility of the quark condensate $\chi(\mu)$ encodes the dynamical
information of the QCD vacuum \cite{Ioffe:1983ju}.

Applying the standard definition for the pion decay constant
\begin{eqnarray}
\langle 0 | j_{\pi}| \pi(p) \rangle = -i \, f_{\pi}  \, \mu_{\pi}(\mu) \,,
\qquad \mu_{\pi}(\mu) \equiv {m_{\pi}^2 \over  m_u(\mu) + m_d(\mu) } \,,
\end{eqnarray}
we can write down the hadronic dispersion relation of $G(p^2, Q^2)$
\begin{eqnarray}
G(p^2, Q^2)  =  {f_{\pi}  \, \mu_{\pi}(\mu) \over m_{\pi}^2  - p^2 - i 0} \,
F^{\rm NLP}_{\gamma^{\ast} \gamma \to \pi^0} (Q^2)
+ \int_{s_0}^{\infty} \,ds \, {\rho^{h}(s, Q^2) \over s-p^2-i 0}  \,.
\end{eqnarray}
The final expression of the LCSR for the hadronic photon correction to the pion-photon form factor
can then be derived by implementing the continuum subtraction and the Borel transformation
with the aid of the parton-hadron duality
\begin{eqnarray}
F^{\rm NLP}_{\gamma^{\ast} \gamma \to \pi^0} (Q^2) &=&
- { \sqrt{2} \, \left ( Q_u^2 - Q_d^2 \right )  \over f_{\pi}  \,\, \mu_{\pi}(\mu)  } \,
\chi(\mu) \, \langle \bar q q \rangle (\mu) \,
\int_{u_0}^1  \, {du \over u } \, {\rm exp} \left [ - {\bar u \, Q^2 - u m_{\pi}^2 \, \over u \, M^2} \right ]
\phi_{\gamma}(u, \mu) \nonumber \\
&& + \, {\cal O}(\alpha_s)  \,,
\end{eqnarray}
with $u_0=Q^2/(s_0+Q^2)$. Employing the power counting scheme for the  sum rule parameters
\begin{eqnarray}
s_0 \sim M^2 \sim {\cal O} (\Lambda^2) \,, \qquad \bar u_0 \sim {\cal O} (\Lambda^2/Q^2) \,,
\end{eqnarray}
we can readily obtain the scaling behaviour of the hadronic photon effect at large $Q^2$
\begin{eqnarray}
{F^{\rm NLP}_{\gamma^{\ast} \gamma \to \pi^0} (Q^2) \over F^{\rm LP}_{\gamma^{\ast} \gamma \to \pi^0} (Q^2)}
\sim {\cal O} \left ( {\Lambda^2 \over Q^2} \right ) \,.
\end{eqnarray}

\subsection{The hadronic photon effect at one loop}

\begin{figure}
\begin{center}
\includegraphics[width=1.0 \columnwidth]{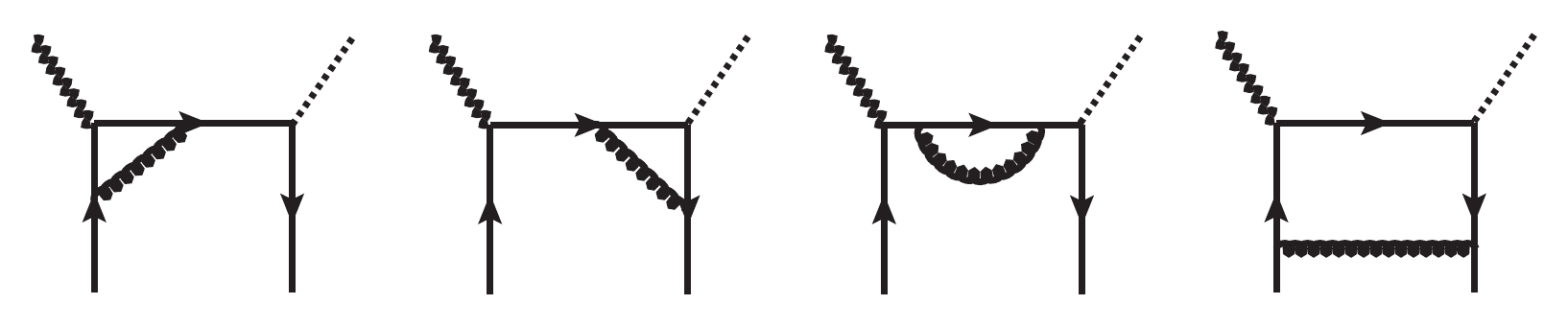} \\
(a) \hspace{3.5 cm} (b)  \hspace{3.5 cm} (c)  \hspace{3.5 cm} (d)
\vspace*{0.1cm}
\caption{Diagrammatical representation of the one-loop contribution to
the QCD amplitude $\widetilde{\Pi}_{\mu}$ (\ref{vacuum to photon correlator: partonic}).
The corresponding symmetric diagrams obtained by exchanging
the electromagnetic current and the pion interpolating current are not shown. }
\label{fig:one-loop-pion-FF-NLP}
\end{center}
\end{figure}

To construct the NLL LCSR for the hadronic photon effect, we first need to establish the one-loop factorization
formula for the correlation function (\ref{vacuum to photon correlator}) at the leading power
in $1/Q^2$. Following the strategy for  demonstrating QCD factorization of    the leading power contribution
presented in Section \ref{sect:leading-power effect},  the perturbative matching coefficient  entering the
factorization formula of the form factor $G(p^2, Q^2)$ can be determined by evaluating the one-loop
diagrams for the QCD amplitude  $\widetilde{\Pi}_{\mu}$ (\ref{vacuum to photon correlator: partonic})
in figure \ref{fig:one-loop-pion-FF-NLP}. We will compute the hard contributions from these
diagrams one-by-one with both the NDR and HV schemes applying the strategy of regions.

The one-loop QCD correction to the electromagnetic vertex diagram displayed in figure
\ref{fig:one-loop-pion-FF-NLP}(a) is obviously free of the $\gamma_5$ ambiguity
and a straightforward calculation yields
\begin{eqnarray}
\widetilde{\Pi}_{\mu}^{(1a)} \big |_{\rm NDR} &=&  \widetilde{\Pi}_{\mu}^{(1a)} \big |_{\rm HV} =
{i \, g_{\rm em} \over 2 \, \sqrt{2}} \, {\bar n \cdot p \over Q^2} \,
 {\alpha_s \, C_F \over 4 \pi}  \, \sum_{q=u\,, d} \eta_q \, Q_q \,
\,  \langle  \widetilde{O}_{1, \mu}(x, x^{\prime})  \rangle^{(0)} \, \ast
\bigg \{  {1 \over x^{\prime} \, r + \bar x^{\prime}} \, {1 \over x^{\prime} \, \bar r} \nonumber \\
&&  \times
\bigg ( \left [ 2 \, \ln \left (x^{\prime} \, r + \bar x^{\prime} \right )  + x^{\prime} \, \bar r \right ] \,
\left [ {1 \over \epsilon} + \ln {\mu^2 \over Q^2} -{1 \over 2}  \, \ln \left (x^{\prime} \, r + \bar x^{\prime} \right )
- {x^{\prime} \, \bar r \over 4 } + {3 \over 2} \right ] \nonumber \\
&& \hspace{0.5 cm} + \, {x^{\prime} \, \bar r  \, (x^{\prime} \, \bar r + 10) \over 4}  \bigg ) \bigg \} + ... \,,
\end{eqnarray}
where the term proportional to $\langle \widetilde{O}_{E, \mu}(x, x^{\prime}) \rangle$ is not shown explicitly.
Due to the appearance of $\gamma_5$ in the pion interpolating current, the one-loop QCD correction to the pion vertex diagram
depends on the $\gamma_5$ prescription employed in the reduction of the Dirac algebra. Computing the hard contribution
from the one-loop diagram  \ref{fig:one-loop-pion-FF-NLP}(b) in both the NDR and HV schemes gives
\begin{eqnarray}
\widetilde{\Pi}_{\mu}^{(1b)} \big |_{\rm NDR} &=&
- {i \, g_{\rm em} \over 2 \, \sqrt{2}} \, {\bar n \cdot p \over Q^2} \,
 {\alpha_s \, C_F \over 4 \pi}  \, \sum_{q=u\,, d} \eta_q \, Q_q \,
\,  \langle  \widetilde{O}_{1, \mu}(x, x^{\prime})  \rangle^{(0)} \, \ast
\bigg \{  {1 \over x^{\prime} \, r + \bar x^{\prime}} \nonumber \\
&&  \times \bigg ( 2 \,  \left [ {r \over  \bar x^{\prime} \, \bar r} \,
\ln \left (  {x^{\prime} \, r + \bar x^{\prime} \over r } \right ) + 1 \right ] \,
\left [ {1 \over \epsilon} + \ln {\mu^2 \over Q^2} -{1 \over 2}  \, \ln r
-  {1 \over 2} \, \ln \left ( x^{\prime} \, r + \bar x^{\prime} \right )
- { \bar x^{\prime} \, \bar r \over 2 \, r }  \right ] \nonumber \\
&& \hspace{0.5 cm} + \, { \bar x^{\prime} \, \bar r \over r } + 2  \bigg ) \bigg \} + ... \,,   \\
\widetilde{\Pi}_{\mu}^{(1b)} \big |_{\rm HV} &=&  \widetilde{\Pi}_{\mu}^{(1b)} \big |_{\rm NDR}  +
 {2 \, \alpha_s \, C_F \over \pi} \, \widetilde{\Pi}_{\mu}^{(0a)} + ...  \,,
\end{eqnarray}
where $\widetilde{\Pi}_{\mu}^{(0a)}$ represents the tree-level contribution to the diagram
\ref{fig:tree-level-pion-FF-NLP}(a) and can be obtained from (\ref{NLP correlator at tree level})
by keeping only the first term in the square bracket.
The self-energy correction to the intermediate hard propagator displayed in figure
\ref{fig:one-loop-pion-FF-NLP}(c) is evidently independent of the $\gamma_5$  prescription
in the $D$-dimensional space and we can readily obtain
\begin{eqnarray}
\widetilde{\Pi}_{\mu}^{(1c)} \big |_{\rm NDR} &=&  \widetilde{\Pi}_{\mu}^{(1c)} \big |_{\rm HV} =
{i \, g_{\rm em} \over 2 \, \sqrt{2}} \, {\bar n \cdot p \over Q^2} \,
 {\alpha_s \, C_F \over 4 \pi}  \, \sum_{q=u\,, d} \eta_q \, Q_q \,
\,  \langle  \widetilde{O}_{1, \mu}(x, x^{\prime})  \rangle^{(0)} \,  \nonumber \\
&&  \, \ast  \, \bigg \{  {1 \over x^{\prime} \, r + \bar x^{\prime}}  \,\,
\left [ {1 \over \epsilon} + \ln {\mu^2 \over Q^2} -  \ln \left (x^{\prime} \, r + \bar x^{\prime} \right )
+ 1 \right ] \bigg \} + ... \,.
\end{eqnarray}
We finally turn to compute the hard contribution from the one-loop box diagram shown in figure
\ref{fig:one-loop-pion-FF-NLP}(d), which depends on the actual prescription  of  $\gamma_5$  adopted in the
calculation of  the corresponding QCD amplitude.
Evaluating the contribution from the box diagram with both the NDR and HV schemes,
we find that the corresponding hard coefficients only contribute at  ${\cal O}(\epsilon)$, vanishing in
four-dimensional space.  Explicitly,
\begin{eqnarray}
\widetilde{\Pi}_{\mu}^{(1d)} \big |_{\rm NDR} &=&  \widetilde{\Pi}_{\mu}^{(1d)} \big |_{\rm HV} = 0  \,.
\end{eqnarray}
Collecting different pieces together, it is straightforward to derive the NLO QCD correction to the four-point amplitude $\widetilde{\Pi}_{\mu}$
\begin{eqnarray}
\widetilde{\Pi}_{\mu}^{(1)} = - {i \, g_{\rm em} \over 2 \, \sqrt{2}} \, {\bar n \cdot p \over Q^2} \,  \sum_{q=u\,, d} \eta_q \, Q_q \,
 \langle  \widetilde{O}_{1, \mu}(x, x^{\prime})  \rangle^{(0)}  \ast  \widetilde{A}_{1, \rm hard}(x^{\prime}) + ... \,,
\end{eqnarray}
where the renormalization prescription dependent hard amplitude $\widetilde{A}_{1, \rm hard}$ is given by
\begin{eqnarray}
\widetilde{A}_{1, \rm hard}(x^{\prime}) \big |_{\rm NDR} &=& {\alpha_s \, C_F \over 4 \pi}  \,
\bigg \{    {1 \over x^{\prime} \, r + \bar x^{\prime}}  \, \bigg [
{2 \over x^{\prime} \, \bar x^{\prime}  \, \bar r} \,
\bigg ( \left ( \left (x^{\prime} \, r - \bar x^{\prime} \right )  \, \ln (x^{\prime} \, r + \bar x^{\prime})
-  x^{\prime} \, r \,  \ln r  \right ) \, \nonumber \\
&& \left ( {1 \over \epsilon} +  \ln {\mu^2 \over Q^2} - {1 \over 2} \, \ln (x^{\prime} \, r + \bar x^{\prime})
-{1 \over 2} \, \ln r \right )  \bigg )  \nonumber
\label{NLP loop amplitude: NDR} \\
&& - {1 \over  x^{\prime} \bar r} \, (\ln r +3) \, \ln (x^{\prime} \, r + \bar x^{\prime})  - 3 \bigg ]
+  (x^{\prime} \leftrightarrow \bar x^{\prime} ) \bigg \} \,, \\
\widetilde{A}_{1, \rm hard}(x^{\prime}) \big |_{\rm HV} &=&  \widetilde{A}_{1, \rm hard}(x^{\prime}) \big |_{\rm NDR}
+  {2 \, \alpha_s \, C_F \over \pi} \,  \widetilde{T}^{(0)}_{1}(x^{\prime}) \,.
\label{NLP loop amplitude: HV}
\end{eqnarray}

Applying the strategy  to implement the IR subtraction for the four-point QCD amplitude $\Pi_{\mu}$
discussed in Section \ref{sect:leading-power effect}, the master formula for the  one-loop
hard coefficient of the physical SCET operator $\widetilde{O}_{1, \mu}$ can be written as
\begin{eqnarray}
\widetilde{T}_1^{(1)} = \widetilde{A}_{1}^{(1)} - \widetilde{T}_1^{(0)} \ast \widetilde{Z}_{11}^{(1)}
+ \widetilde{T}_E^{(0)} \ast  \widetilde{M}_{E1}^{(1) \rm off}
=  \widetilde{A}_{1, \rm hard}^{(1)} + \widetilde{T}_E^{(0)} \ast  \widetilde{M}_{E1}^{(1) \rm off} \,,
\label{subleading power master formula}
\end{eqnarray}
where the bare matrix element $\widetilde{M}_{E1}^{(1) \rm off}$ represents the QCD mixing of the evanescent operator
$\widetilde{O}_{E, \mu}$ into $\widetilde{O}_{1, \mu}$ at one loop.
It is evident that the infrared subtraction term $\widetilde{T}_E^{(0)} \ast  \widetilde{M}_{E1}^{(1) \rm off}$
suffers from the $\gamma_5$ ambiguity in dimensional regularization. The corresponding SCET diagrams at one loop
are in analogy  to that displayed in figure \ref{fig:one-loop-OE-LP}, but with the vertex ``$\otimes$" indicating an insertion
of $\widetilde{O}_{E, \mu}$. Computing these effective diagrams with  dimensional regularization applied to
the UV divergences  and with the IR singularities regularized by the fictitious gluon mass, we find that
$\widetilde{M}_{E1}^{(1) \rm off}$ vanishes at one loop with the NDR scheme of $\gamma_5$ and it receives a
nonvanishing  contribution of ${\cal O}(\epsilon)$ with the HV scheme of $\gamma_5$ from the effective diagram
with a collinear gluon exchange between two external quarks. We are then led to conclude that
\begin{eqnarray}
\widetilde{T}_E^{(0)} \ast  \widetilde{M}_{E1}^{(1) \rm off} \big |_{\rm NDR} =
\widetilde{T}_E^{(0)} \ast  \widetilde{M}_{E1}^{(1) \rm off} \big |_{\rm HV} =0  \,.
\label{subleading power IR subtraction}
\end{eqnarray}
Inserting (\ref{subleading power IR subtraction}) into (\ref{subleading power master formula})
immediately yields
\begin{eqnarray}
\widetilde{T}_1^{(1)}  = \widetilde{A}_{1, \rm hard}^{(1)} \,
\label{final result of the hard function}
\end{eqnarray}
for both the NDR and HV schemes of the $\gamma_5$ matrix, with $\widetilde{A}_{1, \rm hard}^{(1)}$
presented in (\ref{NLP loop amplitude: NDR}) and (\ref{NLP loop amplitude: HV}).
We mention in passing that the $\gamma_5$ scheme dependence of the short-distance
function $\widetilde{T}_1^{(1)}$ will not be cancelled by the one-loop QCD correction
to the twist-2 photon DA defined by the light-cone matrix element of the tensor current,
which is clearly free of the $\gamma_5$ ambiguity in dimensional regularization,
and the $\gamma_5$ ambiguity of  $\widetilde{A}_{1, \rm hard}^{(1)}$ can be traced back to the
renormalization prescription dependence of the QCD amplitude (\ref{vacuum to photon correlator: partonic}) itself.

To preserve the one-loop character of the axial anomaly, an additional finite counterterm
must be introduced \cite{Larin:1993tq}
\begin{eqnarray}
Z_{\rm HV}^{P}(\mu) =   1 -  {2 \, \alpha_s(\mu) \, C_F \over \pi} + {\cal O}(\alpha_s^2) \,,
\end{eqnarray}
when performing the UV renormalization of the pseudoscalar current in the HV scheme.
Making use of  (\ref{tree level hard function}), (\ref{NLP loop amplitude: NDR}),
(\ref{NLP loop amplitude: HV}) and (\ref{final result of the hard function}),
it is  then straightforward to verify that
\begin{eqnarray}
Z_{\rm HV}^{P}(\mu) \left [ \widetilde{T}^{(0)}_{1}(x^{\prime}) +
\widetilde{T}^{(1)}_{1}(x^{\prime}, \mu) \right ]_{\rm HV} =  \left [ \widetilde{T}^{(0)}_{1}(x^{\prime}) +
\widetilde{T}^{(1)}_{1}(x^{\prime}, \mu) \right ]_{\rm NDR} \,
\end{eqnarray}
at one loop, which provides a nontrivial check to justify the obtained one-loop
hard amplitude $\widetilde{T}_{1}$. The NLO factorization formula for the vacuum-to-photon
correlation function can be further derived as follows
\begin{eqnarray}
G(p^2, Q^2) = - {Q_u^2 - Q_d^2 \over \sqrt{2} \, Q^2} \,
\chi(\mu) \, \langle \bar q q \rangle (\mu) \,
\int_0^1 dx \,\left [ \widetilde{T}^{(0)}_{1}(x) +  \widetilde{T}^{(1)}_{1}(x, \mu) \right ]_{\rm NDR} \,
\phi_{\gamma}(x, \mu) + {\cal O}(\alpha_s^2) \,. \hspace{0.5 cm}
\label{NLO factorization of the form factor G}
\end{eqnarray}
With the NLO hard coefficient function $\widetilde{T}^{(1)}_{1}$ at hand,  we can also obtain the one-loop
short-distance function entering the factorization formula of the $H \to J/\psi \, \gamma$ form factor
at leading power in $1/m_H^2$ by taking the $r \to \infty$ limit  of  $\widetilde{T}^{(1)}_{1}$
and by performing the analytical continuation in the variable $p^2$,
which reproduces the expression displayed in (3.17) of \cite{Koenig:2015pha} (see also \cite{Shifman:1980dk,Wang:2013ywc})
computed from an alternative approach precisely.

We are now in a position to demonstrate the factorization-scale independence of
(\ref{NLO factorization of the form factor G}) by employing the RG equation of the leading twist photon DA
\begin{eqnarray}
\mu^2 \, {d \over d \mu^2} \, \left[ \chi(\mu) \, \langle \bar q q \rangle (\mu) \, \phi_{\gamma} (x, \mu) \right ]
= \int_0^1 \, d y \, \widetilde{V}(x, y) \, \left[ \chi(\mu) \, \langle \bar q q \rangle (\mu) \, \phi_{\gamma} (y, \mu) \right ] \,,
\label{RGE of photon  DA}
\end{eqnarray}
with the perturbative expansion of the evolution kernel
\begin{eqnarray}
\widetilde{V}(x, y) = \sum_{n=0}  \, \left ( {\alpha_s \over 4 \pi} \right )^{n+1} \, \widetilde{V}_n(x,y) \,.
\end{eqnarray}
The one-loop renormalization kernel $\widetilde{V}_0(x,y)$ is given by \cite{Lepage:1979zb}
\begin{eqnarray}
\widetilde{V}_0(x, y) = 2\, C_F \, \left [ {\bar  x \over \bar y} \,
 {1 \over x - y}  \, \theta(x-y)
+ {x \over y} \,  {1 \over y - x} \, \theta(y-x)  \right ]_{+} - C_F \, \delta(x-y)  \,.
\end{eqnarray}
Taking into account the factorization scale dependence of $\widetilde{T}^{(1)}_{1}(x, \mu)$,
we can further deduce
\begin{eqnarray}
{d \over d \ln \mu}  \, G(p^2, Q^2) =  - \,{3 \over 2} \, {\alpha_s(\mu) \, C_F \over  \pi}  \,
{Q_u^2 - Q_d^2 \over \sqrt{2} \, Q^2} \,
\chi(\mu) \, \langle \bar q q \rangle (\mu) \,
\int_0^1 dx \, \widetilde{T}^{(0)}_{1}(x)  \,  \phi_{\gamma}(x, \mu) +  {\cal O}(\alpha_s^2) \,.
\hspace{0.5 cm}
\end{eqnarray}
The residual $\mu$-dependence of $G(p^2, Q^2)$ evidently originates from the UV renormalization of the QCD pseudoscalar
current defining the correlation function (\ref{vacuum to photon correlator}).
Taking advantage of the  evolution equation of the QCD renormalization constant for the pseudoscalar current
\cite{Chetyrkin:1997dh,Vermaseren:1997fq}
\begin{eqnarray}
{d \over d \ln \mu}  \, \ln Z_P(\mu)
= \sum_{n=0}  \, \left ( {\alpha_s(\mu) \over 4 \pi} \right )^{n+1} \, \gamma_P^{(n)} \,, \qquad
 \gamma_P^{(0)} = 6 \, C_F \,,
\end{eqnarray}
and distinguishing the renormalization scale of the QCD current from the factorization scale due to the
IR subtraction (see \cite{Wang:2015ndk} for more details), we can then find that the expression for the form factor  $G(p^2, Q^2)$
(\ref{NLO factorization of the form factor G}) is indeed factorization-scale invariant at ${\cal O}(\alpha_s)$.

We proceed to perform the NLL resummation for the parametrically large logarithms in the short-distance
function $\widetilde{T}^{(1)}_{1}$, which can be achieved alternatively by fixing the factorization scale
as $\mu \sim \sqrt{Q^2}$ and by evolving the twist-2 photon DA from the hadronic scale  to that scale.
To this end, we need the two-loop coefficient of the evolution kernel $\widetilde{V}(x, y)$
\cite{Belitsky:1999gu,Mikhailov:2008my,Belitsky:2000yn}
\begin{eqnarray}
\widetilde{V}_1(x, y) = {N_f  \over 2} \, C_F \, \widetilde{V}_N(x, y) +  C_F \, C_A \, \widetilde{V}_G(x, y)
+ C_F^2 \, \widetilde{V}_F(x, y) \,,
\end{eqnarray}
where the explicit expressions of the kernel functions are \cite{Mikhailov:2008my}
\begin{eqnarray}
\widetilde{V}_N(x, y) &=& \left \{  - {4 \over 3}  \, \left [ 2 \, \theta(y-x) \, \widetilde{F}(x, y) \,
\left ( \ln {x \over y}  + {5 \over 3} \right )   \right ]_{+}
+ \left (x \leftrightarrow \bar x \,, y \leftrightarrow \bar y \right ) \right \}
+ {26 \over 9} \, \delta(x -y) \,,  \hspace{0.5 cm}
\\
\widetilde{V}_G(x, y) &=& \left \{  -2 \, \left [ \theta(y-x) \, {x \over y} +
\theta(y- \bar x) \, {\bar x \over  y} \right ]
+ \left (x \leftrightarrow \bar x \,, y \leftrightarrow \bar y \right ) \right \} - \widetilde{H}(x, y) \nonumber \\
&& + \left \{  \left [ 2 \, \theta(y-x)  \, \widetilde{F}(\bar x, \bar y) \,
\left (  {11 \over 3} \, \ln {x \over y} +  {67 \over 9} - {\pi^2 \over 3} \right )  \right ]
+ \left (x \leftrightarrow \bar x \,, y \leftrightarrow \bar y \right )  \right \}   \nonumber \\
&& +  \left [ - {221 \over 18}  -12 \, \zeta(3) + {4 \, \pi^2 \over 3} \right ] \, \delta(x-y)  \,,
\\
\widetilde{V}_F(x, y) &=&  \left \{  4 \, \left [ \theta(y-x) \, {x \over y} +
\theta(y- \bar x) \, {\bar x \over y} \right ]
+ \left (x \leftrightarrow \bar x \,, y \leftrightarrow \bar y \right ) \right \}
+ 2 \, \widetilde{H}(x, y) \nonumber \\
&&  + \, \bigg \{  4 \, \bigg [ \theta(y-x)  \, \bigg ( \widetilde{F}(x, y) \, \ln^2 {x \over y}
+{1 \over y \bar y}  \, \ln x \, \ln \bar x  - {3 \over 2 } \, \widetilde{F}(x, y) \,  \ln {x \over y} \nonumber \\
&& - \left ( \widetilde{F}(x, y)-  \widetilde{F}(\bar x, \bar y) \right ) \, \ln {x \over y} \,
\ln \left (1- {x \over y} \right ) \bigg )  \bigg ]_{+}
+ \left (x \leftrightarrow \bar x \,, y \leftrightarrow \bar y \right )  \bigg \}  \, \nonumber \\
&& + \, 4 \, \left [{11 \over 8}  + 6 \, \zeta(3)  - {2 \pi^2 \over 3} \right ]  \, \delta(x-y)\,,
\end{eqnarray}
with
\begin{eqnarray}
\widetilde{F}(x, y) &=& {x \over y} \, {1 \over y -x} \,, \\
\widetilde{H}(x, y) &=&  -4 \, \bigg  [  \theta(y-x)  \,
\left ( \widetilde{F}(\bar x, \bar y) \, \ln \bar x \, \ln y
- \widetilde{F}(x, y) \, \left [{\rm Li}_2(x) + {\rm Li}_2(\bar y) \right ]
+ {\pi^2 \over 6} \, \widetilde{F}(x, y) \right )\nonumber \\
&&  +  \theta(x -\bar y)  \, \bigg ( \left [ {\rm Li}_2 \left ( 1-{x \over y} \right )
+ {1 \over 2} \, \ln^2 x  \right ]
+  \widetilde{F}(x, y)  \, \left [  {\rm Li}_2(\bar y) - \ln x \, \ln y \right ] \nonumber \\
&& +  \widetilde{F}(\bar x, \bar y) \,  {\rm Li}_2(\bar x)  \bigg )  \bigg ]
+ \left (x \leftrightarrow \bar x \,, y \leftrightarrow \bar y \right )  \,.
\end{eqnarray}
Applying the Gegenbauer expansion of the twist-2 photon DA \cite{Ball:2002ps}
\begin{eqnarray}
\phi_{\gamma}(x ,\mu) = 6 \, x \, \bar x \, \sum_{n=0}^{\infty} \, b_n(\mu) \, C_n^{3/2}(2 x-1)\,
\label{Gegenbauer expansion of photon DA} \,,
\end{eqnarray}
and implementing the conformal consistency relation discussed in \cite{Mueller:1993hg},
the two-loop evolution of the Gegenbauer moment $b_n(\mu_0)$ can be constructed as follows
\begin{eqnarray}
\chi(\mu) \, \langle \bar q q \rangle (\mu) \, b_n(\mu) &=&
E_{T, n}^{\rm NLO}(\mu, \mu_0) \, \chi(\mu_0) \, \langle \bar q q \rangle (\mu_0) \, b_n(\mu_0) \nonumber \\
&& +  {\alpha_s(\mu) \over 4 \pi} \, \sum_{k=0}^{n-2} \,  E_{T, n}^{\rm LO}(\mu, \mu_0) \,
d_{T, n}^{k}(\mu, \mu_0) \, \chi(\mu_0) \, \langle \bar q q \rangle (\mu_0) \, b_n(\mu_0)   \,,
\label{NLL evolution of photon  moments}
\end{eqnarray}
with even $k, n \geq 0$.  The detailed expressions the  evolution functions
 $E_{T, n}^{\rm NLO}$ and $d_{T, n}^{k}$ can be found  in Appendix \ref{app:two-loop evolutions}.
 Combining everything together we arrive at the NLL resummation improved factorization formula
\begin{eqnarray}
G(p^2, Q^2) = - {\left (Q_u^2 - Q_d^2 \right ) \over  \sqrt{2} \,  Q^2} \,
\sum_{n=0} \left [ \chi(\mu) \, \langle \bar q q \rangle (\mu) \, \, b_n(\mu) \right ]
 \, \widetilde{C}_n (Q^2, \mu)\, + {\cal O}(\alpha_s^2)  \,. \hspace{0.5 cm}
\label{NLO factorization of the form factor G with resummation}
\end{eqnarray}
 where the perturbative matching coefficient $ \widetilde{C}_n (Q^2, \mu)$ is defined  by
\begin{eqnarray}
\widetilde{C}_n (Q^2, \mu) =  \int_0^1 \, d x \,
\left [ \widetilde{T}^{(0)}_{1}(x) +  \widetilde{T}^{(1)}_{1}(x, \mu) \right ]_{\rm NDR} \,
\left[ 6 \, x\, \bar x  \, C_n^{3/2}(2 x-1) \right ]  \,.
\label{definition of C function at NLP}
\end{eqnarray}
We will not present the analytical result of  $\widetilde{C}_n (Q^2, \mu)$ by evaluating
the appeared convolution integral explicitly, since the continuum
subtraction needs to be performed for the dispersion representation  of
(\ref{definition of C function at NLP}) in order to construct the desired LCSRs for the hadronic photon
correction to the pion-photon form factor.

Employing the spectral representations of the convolution integrals displayed in Appendix \ref{app:spectral representations},
it is straightforward to derive the dispersion form of the NLL factorization formula
\begin{eqnarray}
G(p^2, Q^2) &=& - { \sqrt{2} \, \left (Q_u^2 - Q_d^2 \right ) \over   Q^2} \,
\chi(\mu) \, \langle \bar q q \rangle (\mu) \,
\int_0^{\infty} \, {ds \over  s -p^2 - i \, 0} \, \nonumber \\
&& \times \left [ \rho^{(0)}(s, Q^2) + {\alpha_s \, C_F \over 4\, \pi} \,  \rho^{(1)}(s, Q^2)  \right ]  \,,
\end{eqnarray}
where we have exploited the symmetric property of the photon DA
$\phi_{\gamma} (x, \mu)=\phi_{\gamma} (\bar x, \mu)$ due to the charge-parity conservation.
The resulting QCD spectral densities $\rho^{(i)}(s, Q^2)$ ($i=0, 1$)  can be written as
\begin{eqnarray}
\rho^{(0)}(s, Q^2) &=& {Q^2 \over Q^2 + s} \,\, \phi_{\gamma} \left ({Q^2 \over Q^2 + s}, \mu \right ) \,, \\
\rho^{(1)}(s, Q^2) &=&   2\, \int_0^1 \,  {  du \over  \bar u} \,
\bigg \{ \theta \left (u - {Q^2 \over Q^2 + s} \right ) \, {Q^2 \over Q^2 + s} \,
\left [  {\bar u - u \over u}  \, \ln \left ({\mu^2 \over u \, s - \bar u \, Q^2} \right )
+ {3 \over 2} \, {\bar u \over u} \right ]  \nonumber \\
&& + \, \ln \left ( {\mu^2 \over s} \right )  \, \left [  {Q^2 \over Q^2 + s}
- {\cal P} {\bar u \, Q^2  \over \bar u \, Q^2 -  u \, s} \right ] \bigg \} \,
\phi_{\gamma} \left (u, \mu \right )  \nonumber \\
&& + \, {Q^2 \over Q^2+ s} \, \int_0^1 \, d u \,\, \theta \left (u- {Q^2 \over Q^2 + s} \right ) \,
\bigg \{ 2 \, \ln \left ( {u \, s - \bar u \, Q^2 \over  Q^2 } \right ) \,
\bigg  [  \ln \left ( {\mu^2 \over u \, s - \bar u \, Q^2 } \right )  \nonumber \\
&& + \ln \left ( {\mu^2 \over Q^2} \right )  +  {3 \over 2} \bigg ]
- \ln^2 \left ( {\mu^2 \over Q^2} \right ) +  \ln^2 \left ( {\mu^2  \over  s} \right )
- {\pi^2 \over 3} + 3 \bigg \} \, {d \over d u} \,   \phi_{\gamma}(u, \mu)   \,,
\end{eqnarray}
where ${\cal P}$ indicates the principle-value prescription. The NLL LCSRs for the subleading
power contribution to the $\pi^0 \gamma^{\ast} \gamma$ form factor can be further derived as
\begin{eqnarray}
F^{\rm NLP}_{\gamma^{\ast} \gamma \to \pi^0} (Q^2) &=&
- { \sqrt{2} \, \left ( Q_u^2 - Q_d^2 \right ) \over f_{\pi}  \,\, \mu_{\pi}(\mu) \,\, Q^2} \,
\chi(\mu) \, \langle \bar q q \rangle (\mu) \,
\int_0^{s_0}  \, ds \, {\rm exp} \left[ - {s - m_{\pi}^2  \over M^2} \right ] \nonumber \\
&& \times \left [ \rho^{(0)}(s, Q^2) + {\alpha_s \, C_F \over 4\, \pi} \,  \rho^{(1)}(s, Q^2)  \right ]
+ {\cal O}(\alpha_s^2)  \,.
\label{NLL resummation for the NLP effect}
\end{eqnarray}

Collecting different contributions together, we now present the final expression for the pion-photon form factor
including  the twist-4 correction computed in \cite{Agaev:2010aq,Khodjamirian:1997tk}
\begin{eqnarray}
F_{\gamma^{\ast} \gamma \to \pi^0} (Q^2)=F^{\rm LP}_{\gamma^{\ast} \gamma \to \pi^0} (Q^2)
+ F^{\rm NLP}_{\gamma^{\ast} \gamma \to \pi^0} (Q^2) +  F^{\rm tw-4}_{\gamma^{\ast} \gamma \to \pi^0} (Q^2) \,,
\label{final result of the pion-photon FF}
\end{eqnarray}
where the manifest expressions of $F^{\rm LP}_{\gamma^{\ast} \gamma \to \pi^0}$ and
$F^{\rm NLP}_{\gamma^{\ast} \gamma \to \pi^0}$ are displayed in (\ref{NLL resummation for the LP effect})
and (\ref{NLL resummation for the NLP effect}), respectively. The obtained factorization formula of
$F^{\rm tw-4}_{\gamma^{\ast} \gamma \to \pi^0}$ from both the two-particle and the three-particle
pion DAs at tree level reads
\begin{eqnarray}
F^{\rm tw-4}_{\gamma^{\ast} \gamma \to \pi^0} (Q^2) =
- { \sqrt{2} \, \left ( Q_u^2 - Q_d^2 \right ) \over Q^4  }\,
\int_0^1 d x \, { \mathbb{F}_{\pi}(x, \mu) \over x^2 } + {\cal O}(\alpha_s) \,,
\end{eqnarray}
where the definition of the twist-4 pion  DA $\mathbb{F}_{\pi}$ can be found in (38) of \cite{Agaev:2010aq}
and keeping only the leading conformal spin (i.e., ``S"-wave) contribution we obtain
\begin{eqnarray}
\mathbb{F}_{\pi}(x, \mu) = {80 \over 3} \, \delta_{\pi}^2 (\mu) \, x^2 \, (1-x)^2    \,.
\end{eqnarray}
The nonperturbative parameter $\delta_{\pi}^2$ is defined by the local QCD matrix element
\begin{eqnarray}
\langle 0 |g_s \, \bar q \, \tilde{G}_{\mu \nu} \, \gamma^{\nu}  \, q | \pi(p)\rangle
= i \, f_{\pi} \,  \delta_{\pi}^2 (\mu) \, p_{\mu}    \,,
\end{eqnarray}
with the renormalization-scale evolution at one loop
\begin{eqnarray}
\delta_{\pi}^2 (\mu) = \left [ {\alpha_s(\mu) \over \alpha_s(\mu_0)} \right ]^{32 \over 9  \beta_0}\,
 \delta_{\pi}^2 (\mu_0)  \,.
\end{eqnarray}
Several comments on the general structure  of the $\pi^0 \gamma^{\ast} \gamma$ form factor
 (\ref{final result of the pion-photon FF})  are in order.

\begin{itemize}
\item{It is apparent that the twist-four correction to the $\pi^0 \gamma^{\ast} \gamma$ form factor
is suppressed by a factor of $\delta_{\pi}^2/Q^2$ compared with that of
the leading twist contribution. Such subleading power contribution turns out to be numerically significant
at $Q^2 \leq 5 \, {\rm GeV}^2$ due to the large prefactor ``$80/3$" entering the asymptotic expression of  $\mathbb{F}_{\pi}(x, \mu)$,
however, it is still far from sufficient to generate the scaling violation at $Q^2 \sim 40 \, {\rm GeV}^2$ indicated by
the BaBar measurement \cite{Aubert:2009mc}.  Furthermore, it is of high interest to compute the NLO correction
to the twist-four contribution in order to develop a better understanding of factorization properties
of the high twist effects, where the infrared subtraction for constructing the factorization formula
is complicated by the mixing of different twist-four pion DAs  under the QCD renormalization.}

\item{The twist-six correction to the pion-photon form factor computed from the dispersion approach \cite{Agaev:2010aq}
is partially absorbed into the hadronic photon effect $F^{\rm NLP}_{\gamma^{\ast} \gamma \to \pi^0}(Q^2)$ displayed
in (\ref{NLL resummation for the NLP effect}). The precise correspondence of distinct contributions in two frameworks
cannot be established without identifying the operator definitions of the ``soft" corrections in \cite{Agaev:2010aq},
which originate from the nonperturbative modification of the QCD spectral density appeared in the dispersion form of
the $\pi^0 \gamma^{\ast} \gamma^{\ast}$ form factor. }

\item{The subleading power corrections from the yet higher twist pion/photon DAs, which are not taken into account in this work,
are conjectured to be suppressed by {\it only}  one power of $\Lambda^2/Q^2$ due to the absent correspondence
between the  twist counting and the large-momentum expansion \cite{Agaev:2010aq}. A manifest calculation of the two-particle
and three-particle corrections to the pion-photon form factor from the twist-three and twist-four photon DAs based upon
the LCSR approach is in demand to verify  this  interesting hypothesis. }

\item{We do not include the NNLO QCD correction to the leading power contribution in the large $\beta_0$ approximation
\cite{Melic:2002ij} on account of the absence of a complete NNLO contribution, which also necessitates
the three-loop evolution equation of the twist-two pion DA \cite{Braun:2017cih} to obtain the factorization formula
at the next-to-next-to-leading-logarithmic accuracy. A recent discussion of the NNLO radiative corrections
in the framework of the dispersion approach can be found in \cite{Mikhailov:2016klg}.}

\end{itemize}

\section{Numerical analysis}
\label{sect:numerical analysis}

We are now ready to explore the phenomenological consequences of the hadronic photon correction
to the pion-photon form factor applying the master formula (\ref{final result of the pion-photon FF}).
In doing so, we will first need to specify the non-perturbative models for the twist-2 pion DA,
the magnetic susceptibility $\chi(\mu)$, the Gegenbauer moments of the photon DA, and to determine the
``internal" sum rule parameters entering the expression (\ref{NLL resummation for the NLP effect}).

\subsection{Theory input parameters}

The fundamental ingredients entering the NLL factorization formula of the leading power contribution
are the  Gegenbauer moments of the twist-2 pion DA. Tremendous efforts have been devoted to the determinations
of the lowest moment $a_2(\mu)$ from the direct calculations with the QCD sum rules pioneered by
Chernyak and Zhitnitsky (CZ) \cite{Chernyak:1981zz} and with the lattice simulations,
and from the indirect calculations by matching the LCSR predictions with the experimental data.
To quantify the systematic uncertainty from the  Gegenbauer moments, we will consider
the following four  models for the leading twist pion DA
\begin{eqnarray}
a_2(1.0 \, {\rm GeV})  &=& 0.21^{+0.07}_{-0.06}  \,,  \qquad
a_4(1.0 \, {\rm GeV})  = - \left ( 0.15^{+0.10}_{-0.09}  \right ) \,,
\qquad ({\rm BMS}) \,;  \nonumber  \\
a_2(1.0 \, {\rm GeV})  &=& 0.17 \pm 0.08   \,, \qquad
a_4(1.0 \, {\rm GeV})  = 0.06 \pm 0.10  \,, \qquad ({\rm KMOW})  \,;  \nonumber \\
a_n(1.0 \, {\rm GeV})  &=&  \ {2 \, n+3 \over 3 \, \pi}  \, \left ({\Gamma [(n+1)/2] \over \Gamma [(n+4)/2]} \right )^2 \,,
\qquad \hspace{3.0 cm} ({\rm Hol.}) \,;  \nonumber \\
a_2(1.0 \, {\rm GeV})  &=& 0.5 \,,  \qquad a_{n>2}(1.0 \, {\rm GeV})= 0 \,, \qquad
 \hspace{2.9 cm}  ({\rm CZ}) \,.
 \label{values of pion moments}
\end{eqnarray}

The obtained Gegenbauer coefficients in the Bakulev-Mikhailov-Stefanis (BMS) model \cite{Mikhailov:2016klg,Bakulev:2001pa} are computed from
the QCD sum rules with non-local condensates absorbing the high-order terms in the operator-product-expansion (OPE) partially
(see, however, \cite{Chernyak:2006ms}). The first and  second nontrivial  Gegenbauer moments of the KMOW model \cite{Khodjamirian:2011ub}
are determined by comparing the LCSR predictions for the pion electromagnetic form factor, including the NLO correction to the twist-2 effect
and the subleading terms up to twist-6, with the intermediate-$Q^2$ data from the JLab experiment.
The holographic model of the twist-2 pion DA \cite{Brodsky:2007hb}
\begin{eqnarray}
\phi_{\pi}^{\rm Hol}(x,  \mu_0)={8 \over \pi} \,  \sqrt{ x \, (1-x)} \,
\end{eqnarray}
is motivated by the correspondence between the string theory in the five-dimensional
anti-de Sitter space and conformal field theories in the physical space-time
(see also \cite{Cloet:2013tta} for a similar  end-point behaviour of the pion DA) and implementing the
Gegenbauer expansion of $\phi_{\pi}^{\rm Hol}(x,  \mu_0)$ leads to the expression of $a_n$ displayed in (\ref{values of pion moments}).
For the phenomenological analysis of the  $\pi^0 \gamma^{\ast} \gamma$ form factor, we will truncate the expansion of the
``holographic" model at $n=12$, which was demonstrated to be a good approximation in \cite{Agaev:2010aq}.
It needs to point our that the values of the second Gegenbauer coefficient in the first three models of  (\ref{values of pion moments})
are in line with the recent lattice determinations  \cite{Braun:2015axa} within the theory uncertainties
and the CZ model is introduced for the illustration purpose to understand the model dependence of the
predictions for the pion-photon form factor.

The normalization parameter  for the twist-four pion DAs will be taken as
$\delta_{\pi}^2(1 \, {\rm GeV})=(0.2 \pm 0.04) \, {\rm GeV^2}$ computed from the QCD sum rules \cite{Novikov:1983jt}
(see also \cite{Ball:2006wn}). We further adopt the value of the quark condensate density
$\langle \bar q q \rangle (1 \, {\rm GeV})=- \left (256^{+14}_{-16} \, {\rm MeV} \right )^3$ determined in \cite{Khodjamirian:2011ub}.
A key nonperturbative quantity appearing in the twist-2 photon DA is the magnetic susceptibility of the quark condensate
$\chi(\mu)$ describing a response of the QCD vacuum in the presence of an external photon field.
Different QCD-based approaches have been proposed to evaluate $\chi(\mu)$
(see, e.g., \cite{Ball:2002ps,Rohrwild:2007yt,Vainshtein:2002nv}) with the aid of the resonance information
from the experimental data and the interval $\chi(1 \, {\rm GeV})= (3.15 \pm 0.3) \, {\rm GeV^{-2}}$  \cite{Ball:2002ps}
will be employed in the numerical calculations. In contrast, our understanding of the higher Gegenbauer moments of
the leading twist photon DA is rather limited, even for the leading non-asymptotic correction due to $b_2(\mu_0)$.
The available information of the second Gegenbauer coefficient  mainly comes from the  QCD sum rules constructed from
the correlation function with a light-ray tensor operator and a local vector current, which unfortunately give rise to
the theory predictions sensitive to the choice of the input parameters. The crude estimate
$b_2(1 \, {\rm GeV})=0.07 \pm 0.07$ from \cite{Ball:2006wn} will be used in our numerical analysis and an independent
determination  from the lattice QCD calculation will be very welcome in the future.

A natural choice of the factorization scale in the leading power factorization formula
(\ref{NLL resummation for the LP effect}) is $\mu^2= \langle \bar x\rangle  \, Q^2$
with $1/4  \leq \langle \bar x\rangle \leq 3/4$  corresponding to  the  characteristic
 virtuality  of the intermediate  quark displayed in figure \ref{fig:tree-level-pion-FF-LP}(a),
 and it will be frozen at $\mu=1 \, {\rm GeV}$ for $\langle \bar x\rangle  \, Q^2 < 1 \, {\rm GeV^2}$
at low $Q^2$ in order not to run into the nonperturbative QCD regime (see \cite{Melic:2002ij} for the discussion
about the BLM proposal). Along similar lines,  the factorization scale entering the NLL LCSRs for
the hadronic photon effect (\ref{NLL resummation for the NLP effect}) will be taken as
$\mu^2= \langle x \rangle  \, M^2 + \langle \bar x \rangle  \, Q^2$ as widely
employed in the sum rule calculations \cite{Agaev:2010aq}.

Finally, the determination of the Borel mass $M^2$ and the threshold parameter $s_0$ can be achieved
by applying the standard strategies described in \cite{Wang:2015vgv,Wang:2017jow}, and we can readily obtain
\begin{eqnarray}
M^2= (1.25 \pm 0.25) \, {\rm GeV^2}   \,, \qquad
s_0= (0.70 \pm 0.05) \, {\rm GeV^2}   \,,
\end{eqnarray}
in agreement with the intervals adopted in \cite{Khodjamirian:2006st}.

\subsection{Predictions for the $\pi^0 \gamma^{\ast} \gamma$ form factor}

\begin{figure}
\begin{center}
\includegraphics[width=0.46 \columnwidth]{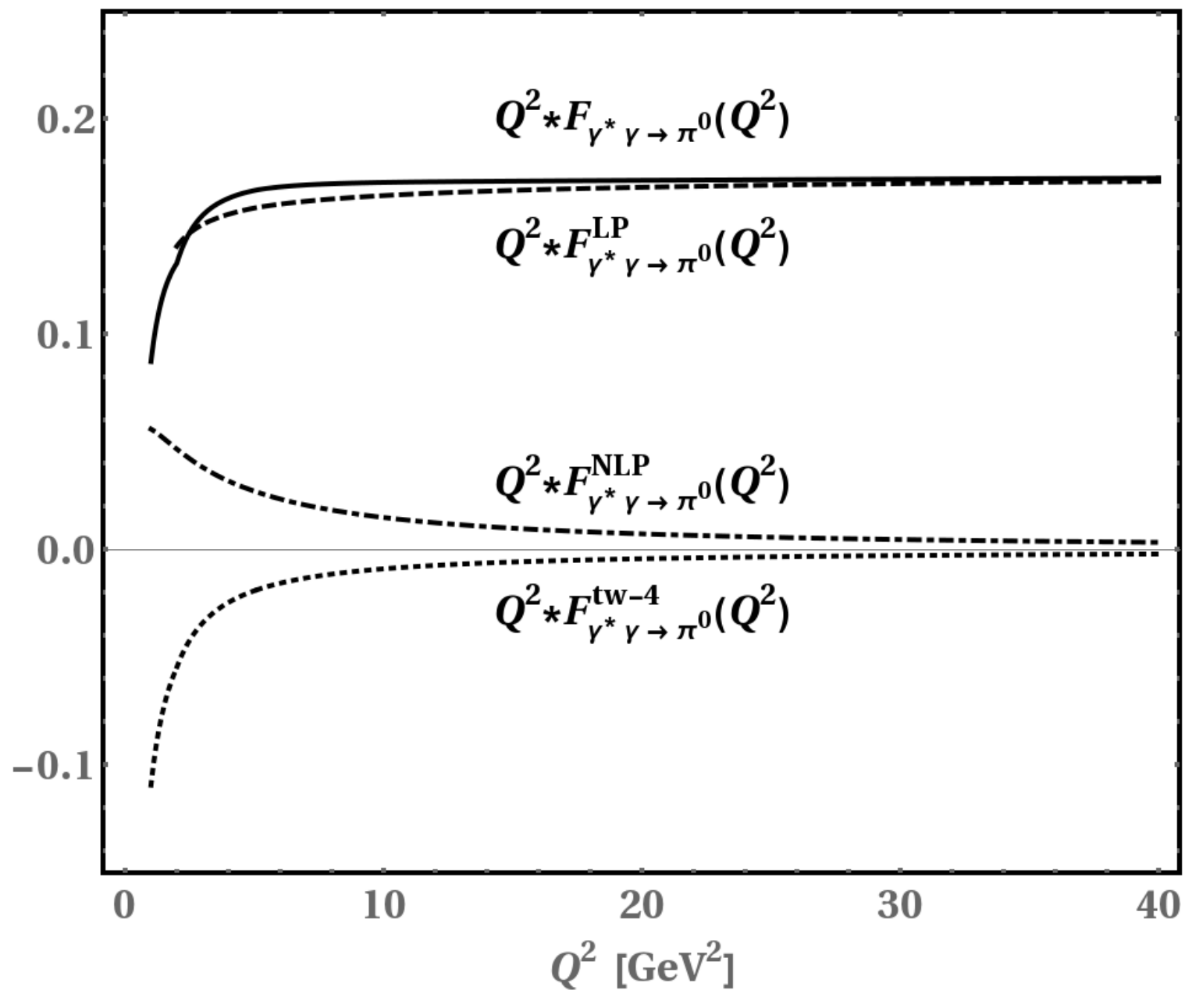} \hspace{1.0 cm}
\includegraphics[width=0.46 \columnwidth]{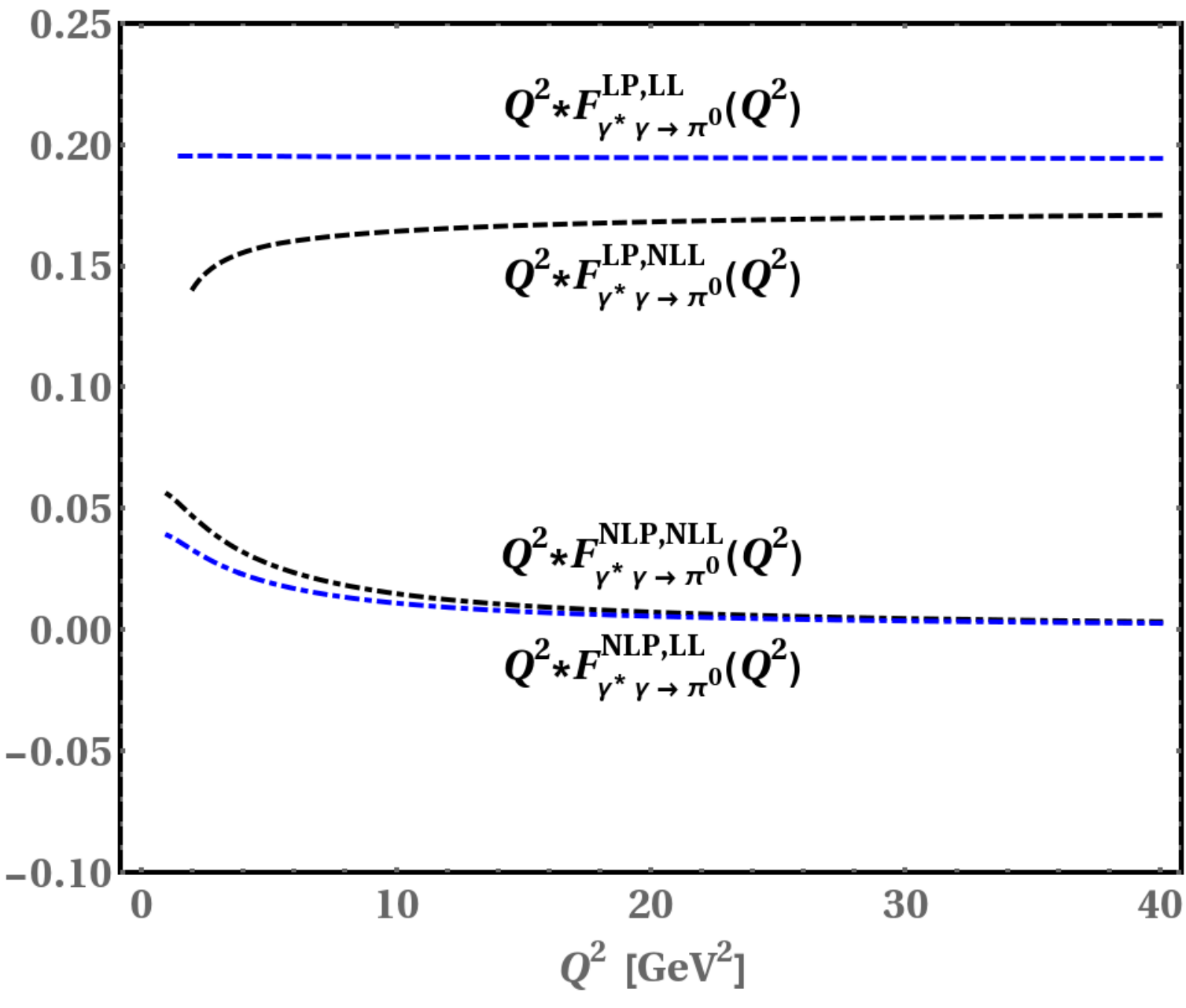}
\vspace*{0.1cm}
\caption{{\bf Left}: Distinct contributions to the $\pi^0 \gamma^{\ast} \gamma$ form factor
from the twist-two pion DA (``{\rm LP}") with the BMS model at NLL, from the
hadronic photon effect (``{\rm NLP}") at NLL and from the twist-four pion DAs
(``{\rm tw-4}") at LO. The solid curve is obtained by adding up the above-mentioned three pieces together
with the central inputs. {\bf Right}: Dependence of perturbative QCD corrections to the leading power
contribution and to the hadronic photon effect with the BMS model, at LL and NLL accuracy, on the  momentum transfer accessible
at the BaBar and Belle experiments. }
\label{fig:form-factor-breakdown}
\end{center}
\end{figure}

Now we will turn to investigate the phenomenological significance of distinct terms contributing
to the pion-photon form factor. Taking the BMS model for the twist-two pion DA as an example,
it is evident from  figure \ref{fig:form-factor-breakdown} that the twist-four correction and the hadronic photon contribution generate
the destructive and constructive interference with the leading power effect (a similar observation for the high twist
corrections already made in \cite{Agaev:2010aq}) and there appears to be
a strong cancellation between these two mechanisms in the whole $Q^2 \leq 40 \, {\rm GeV^2}$ region.
However, both subleading power effects become rapidly suppressed with the growing of the momentum transfer squared
in contrast to the numerically sizeable soft power correction estimated from the dispersion approach \cite{Agaev:2010aq}.
Such discrepancy may be ascribed to the very definition of the ``soft" effect in the formalism of \cite{Khodjamirian:1997tk},
roughly corresponding to  the $\rho$-resonance contribution to the $\pi^0 \gamma^{\ast} \gamma$ form factor
with the parton-hadron duality approximation,  which has no transparent counterpart in the framework of
 perturbative QCD factorization. In addition, the NLL radiative corrections are observed to give rise to
 approximately ${\cal O} (15 \, \% )$ (almost $Q^2$-independent) shift to the LL predictions for
 both the leading power contribution and the hadronic photon effect.

\begin{figure}
\begin{center}
\includegraphics[width=0.46 \columnwidth]{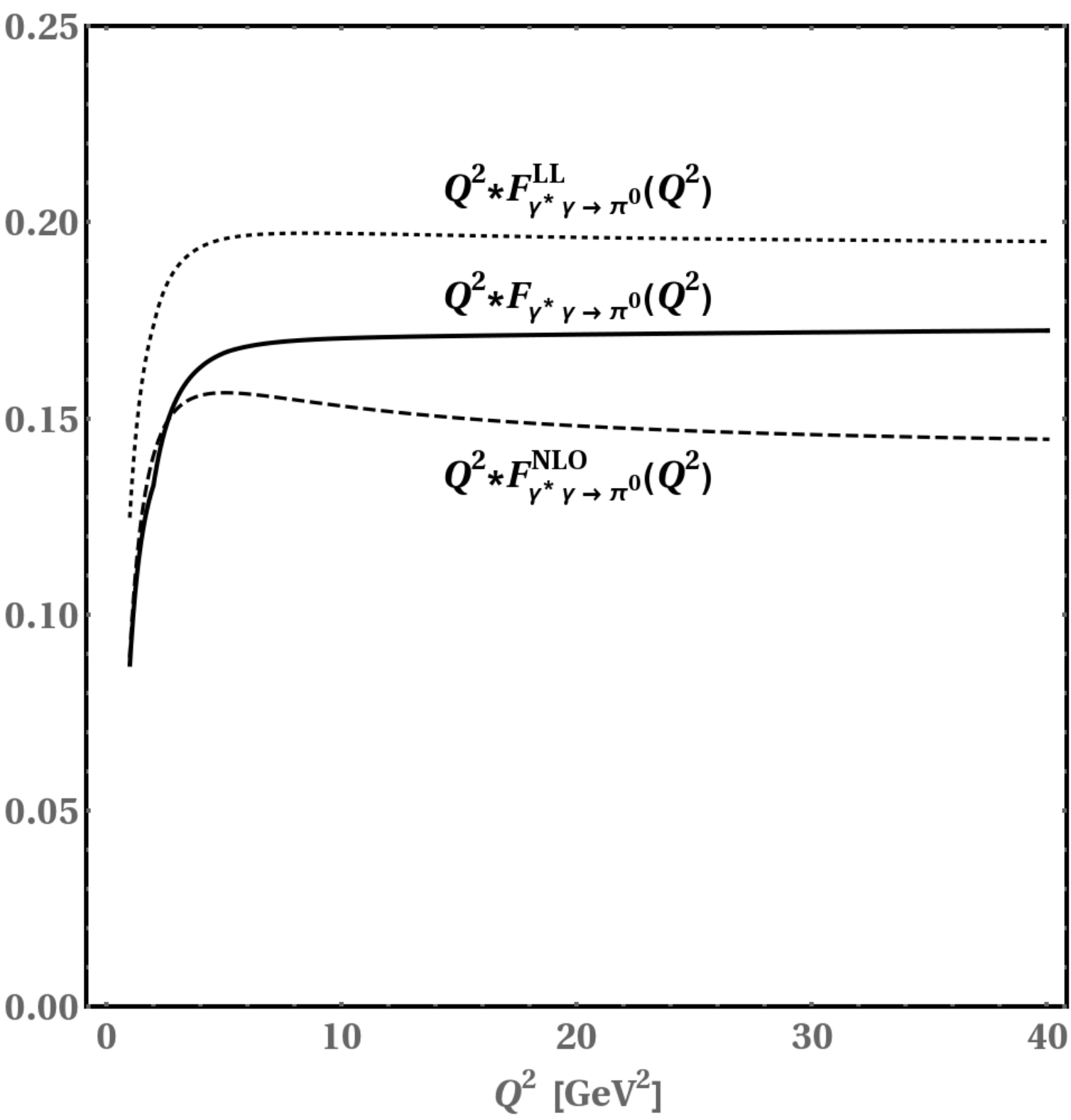} \hspace{1.0 cm}
\includegraphics[width=0.46 \columnwidth]{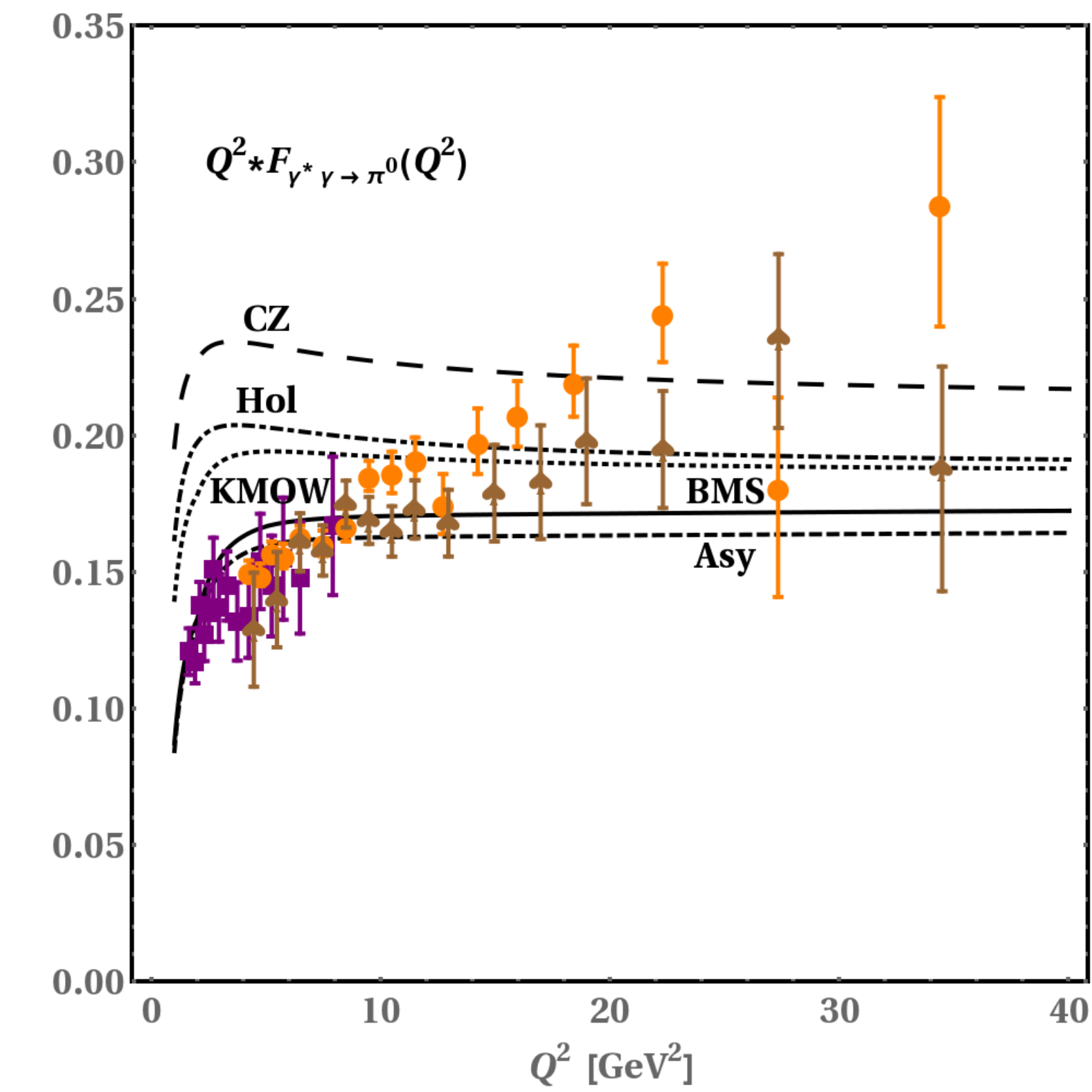}
\vspace*{0.1cm}
\caption{{\bf Left}: The $Q^2$ dependence of the LL, NLO and NLL contributions
to the $\pi^0 \gamma^{\ast} \gamma$ form factor with the BMS model.
{\bf Right}: Theory predictions for the  pion-photon form factor with different models of the
twist-two pion DA presented in (\ref{values of pion moments}). The experimental data are taken from
CLEO \cite{Gronberg:1997fj} (purple squares), BaBar \cite{Aubert:2009mc} (orange circles)
and Belle \cite{Uehara:2012ag} (brown spades).}
\label{fig:form-factor-resum-and-model-dependence}
\end{center}
\end{figure}

\begin{figure}
\begin{center}
\includegraphics[width=0.46 \columnwidth]{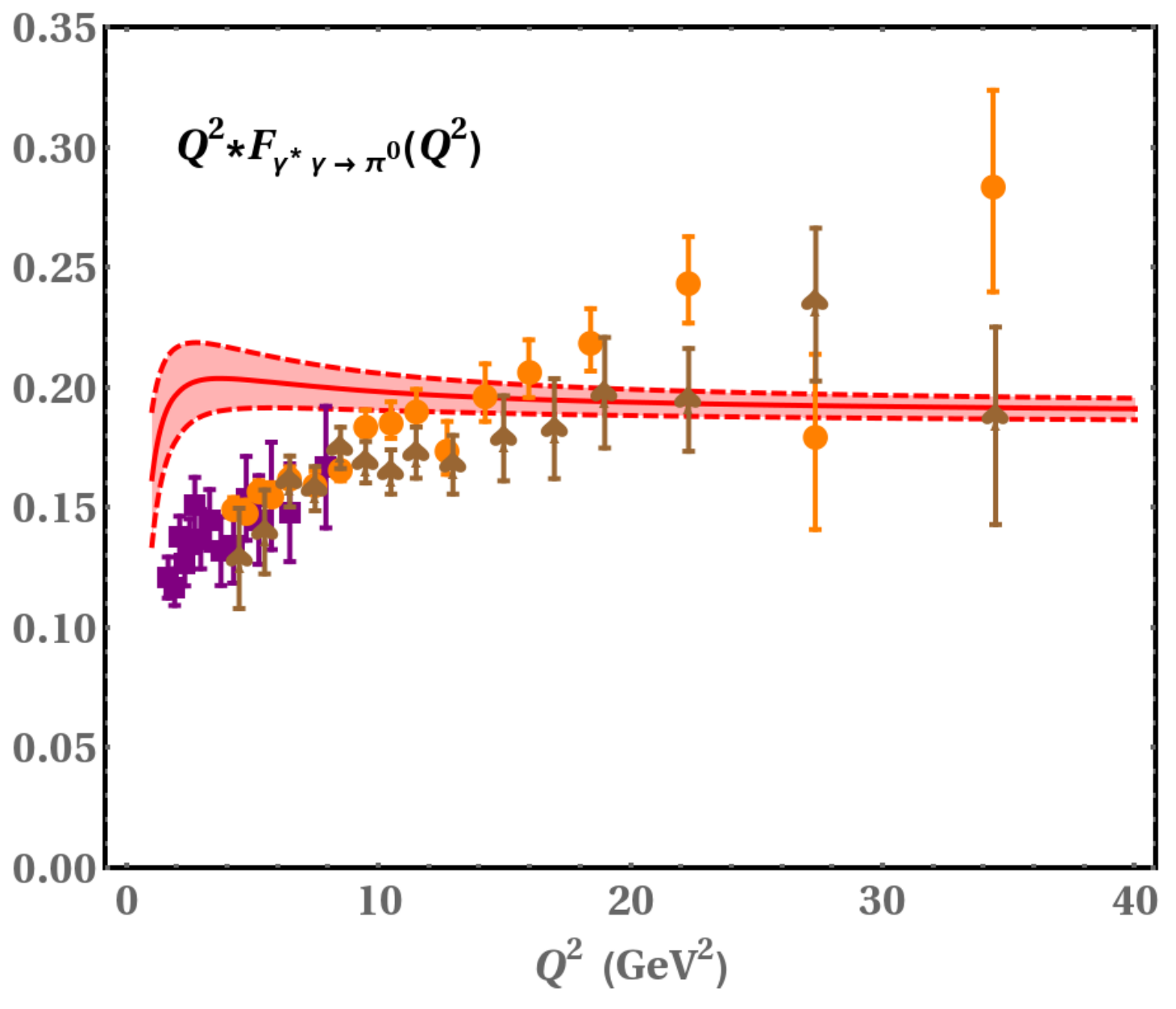} \hspace{1.0 cm}
\includegraphics[width=0.46 \columnwidth]{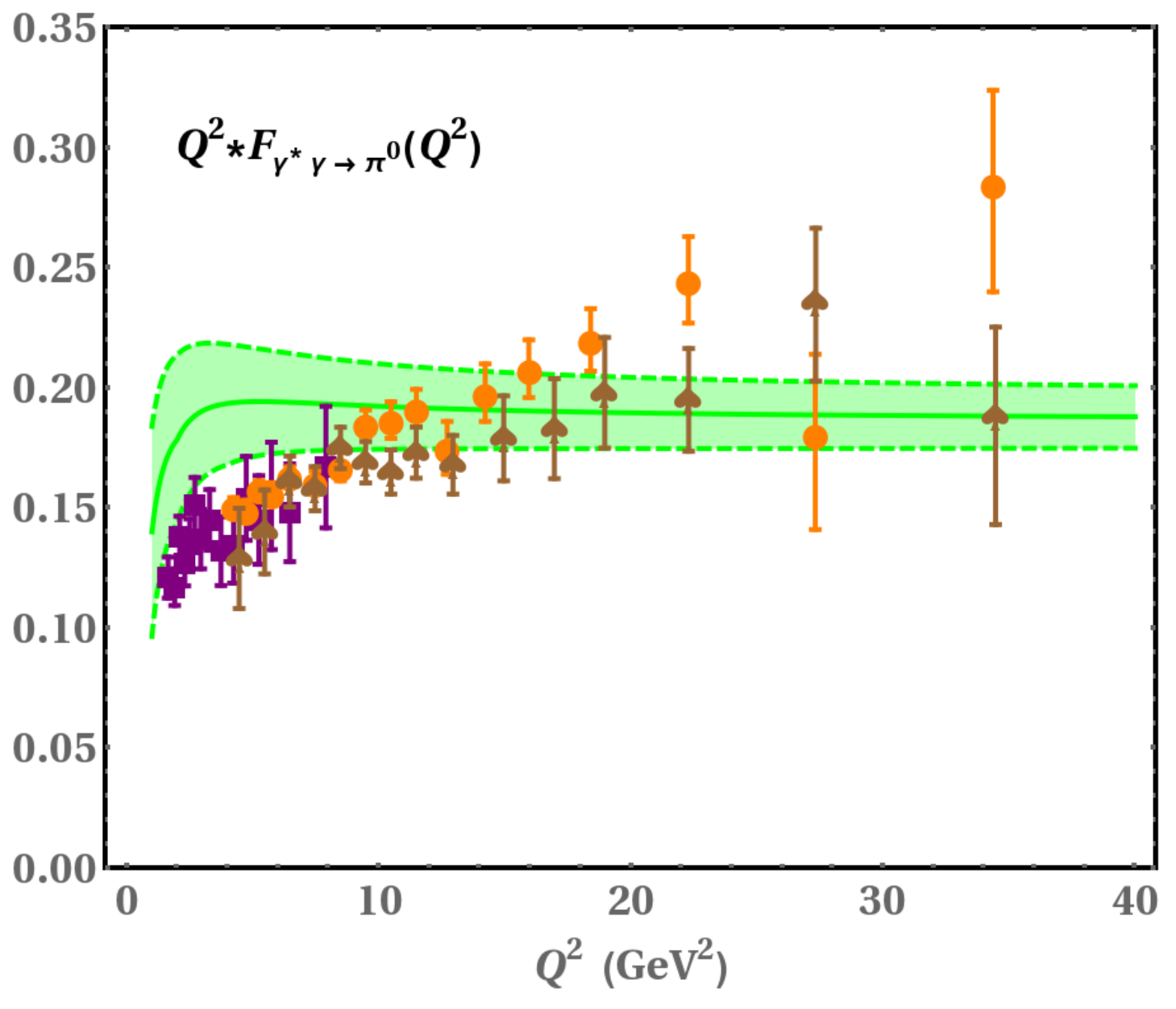} \\
(a)  \hspace{7 cm}    (b) \\
\vspace{0.50 cm}
\includegraphics[width=0.46 \columnwidth]{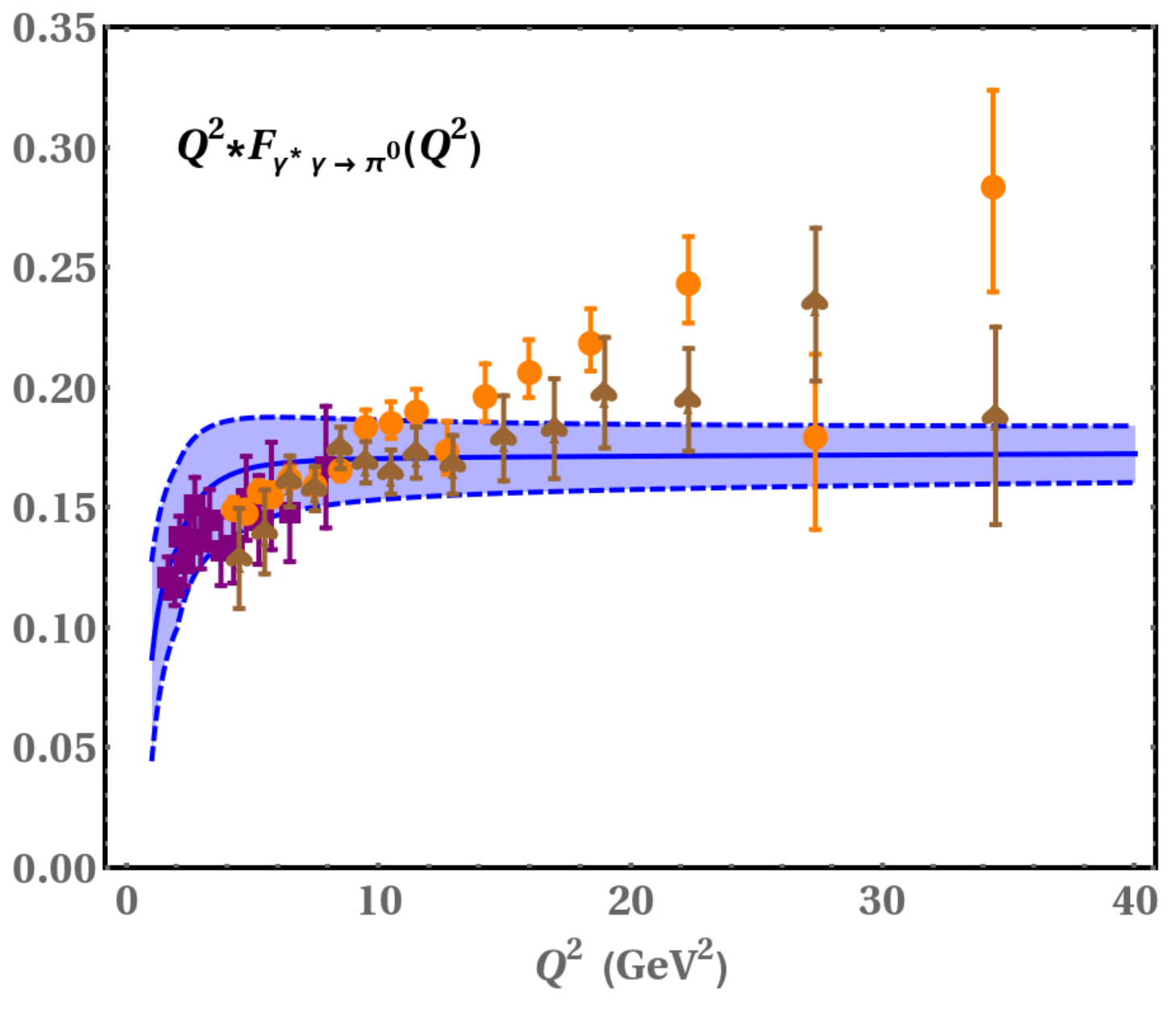} \\
(c) \\
\vspace*{0.1cm}
\caption{The $Q^2$ dependence of the $\pi^0 \gamma^{\ast} \gamma$ form factor computed from
 (\ref{final result of the pion-photon FF}) with
(a) the holographic  model,  (b) the KMOW model,  and (c)  the BMS model.
The shaded regions represent the combined theory uncertainties obtained by adding  the separate errors
in quadrature. The experimental data points from  CLEO \cite{Gronberg:1997fj} (purple squares), 
BaBar \cite{Aubert:2009mc} (orange circles) and Belle \cite{Uehara:2012ag} (brown spades)
are also displayed here.}
\label{fig:form-factor-uncertainty}
\end{center}
\end{figure}

To understand the phenomenological impact of the QCD resummation for the large logarithms
appearing in the factorization formula for the leading power contribution
and in the LCSRs for the hadronic photon correction,
we further present in figure \ref{fig:form-factor-resum-and-model-dependence} our predictions
for the  $\pi^0 \gamma^{\ast} \gamma$ form factor, at LL, NLO and NLL accuracy, with the BMS model.
The NLO QCD corrections are found to induce ${\cal O}\, (25 \, \%)$ reduction of the tree-level results
at $10 \, {\rm GeV^2} \leq  Q^2 \leq 40 \, {\rm GeV^2}$, however, the NLL resummation effect will
enhance the  NLO predictions by an amount of  approximately ${\cal O}\, (10 \, \%)$,
in accordance with the pattern for the perturbative QCD corrections observed in \cite{Wang:2015vgv,Wang:2016qii}.
Inspecting the model dependence of pion-photon form factor on the leading twist pion DA displayed in
figure \ref{fig:form-factor-resum-and-model-dependence} implies that the theory predictions with both the
holographic and KMOW models  can reasonably balance the BaBar and Belle data at high $Q^2$ without resorting to
the ``exotic" end-point behaviour as advocated in \cite{Radyushkin:2009zg,Polyakov:2009je}.
In fact, we have checked that the predicted  $\pi^0 \gamma^{\ast} \gamma$ form factor with the flat pion DA
will overshoot both the BaBar and Belle data, in most $Q^2$ region of interest, at least in our framework.
Given the fact that the end-point behaviour of the twist-two pion DA in the holographic  model differs from
the standard postulation, motivated by the conformal expansion analysis, as employed for the KMOW model,
we conclude that the local information of the pion DA cannot be extracted from the experimental measurements
of the pion-photon form factor even in the leading power approximation.
It needs further to point out that our predictions with the  holographic and KMOW models
do not match the experimental data at $2 \, {\rm GeV^2}\leq  Q^2 \leq 8 \, {\rm GeV^2}$ well,
where the power suppressed contributions from the yet higher-twist pion and photon DAs will become more pronounced
and actually  the large-momentum expansion applied for the construction of the factorization formula also becomes questionable.
By contrast, the theory predictions from the dispersion approach \cite{Agaev:2010aq,Agaev:2012tm}
can result in a satisfactory description of the BaBar and Belle
data in the whole $Q^2$ region by introducing the nonperturbative modification of the  QCD spectral density function.
Moreover, it becomes apparent that the computed pion-photon form factor with the BMS model and the asymptotic pion DA
are less favorable by the experimental measurements at high $Q^2$, albeit with the reasonable agreement achieved
at low $Q^2$. Also,  confronting the theory predictions from the CZ model with the BaBar and Belle data indicates
a large value of  the second Gegenbauber moment $a_2(\mu_0)$ is not favored, in agreement with the recent lattice QCD
calculations \cite{Braun:2015axa,Bali:2017ude}.

We present our final predictions for the $\pi^0 \gamma^{\ast} \gamma$ form factor from the expression
(\ref{final result of the pion-photon FF}) with three different models of the twist-two pion DA
in figure \ref{fig:form-factor-uncertainty}, including the theory uncertainties
due to the variations of the input parameters discussed before.
We already assigned $20 \%$ uncertainty for the first six nontrivial Gegenbauer coefficients of
the holographic model in the numerical estimation for the illustration purpose.
It turns out that the dominant theory uncertainties originate from the shape parameters of the leading twist pion
and photon DAs instead of the variations of the factorization scales.
Precision determinations of the  higher Gegenbauer coefficients for both two DAs along the lines of
\cite{Braun:2015axa,Bali:2017ude} will be essential to pin down the presently sizeable theory uncertainty
in order to meet the challenge of the (potentially) more accurate experimental measurements at the BEPCII collider
\cite{Unverzagt:2012zz} and the SuperKEKB accelerator.

\section{Conclusion}
\label{sect:conclusion}

Applying the standard OPE technique with the evanescent operator(s) we revisited the demonstration
of QCD factorization for the pion-photon  transition form factor at leading power in $1/Q^2$
with both the NDR and HV schemes for $\gamma_5$ in the $D$-dimensional space.
It has been shown explicitly at one loop  that the renormalization scheme dependence of the short-distance
matching coefficient and the twist-two pion DA are cancelled out precisely rendering the $\gamma_5$-prescription
independence of the factorization formula for the leading power contribution to $F_{\gamma^{\ast} \gamma \to \pi^0} (Q^2)$.
This can be readily understood from the fact that the QCD matrix element defined by two
electromagnetic currents are free of the $\gamma_5$ ambiguity and the renormalization scheme dependence of the
hard function arises from the infrared subtraction term completely.
In the same vein, we established QCD factorization of the desired correlation function at one loop
for the construction of the LCSRs for the hadronic photon contribution to the pion-photon form factor.
By contrast, the corresponding QCD matrix element defined with an interpolating current for the pion and
an electromagnetic current suffers from the $\gamma_5$ ambiguity and the leading twist photon DA is independent
of the $\gamma_5$ prescription in dimensional regularization.
The finite renormalization term introduced in the HV scheme to restore the appropriate Ward-Takahashi identities
was found to provide the very transformation function to construct the hard matching coefficient in the NDR scheme.
The NLL resummation of the parametrically large logarithms was also implemented by solving the relevant two-loop
evolution equations in momentum space.

Taking into account the leading power contribution and the hadronic photon effect at NLL and
the twist-four pion DA correction at tree level, we further explored the phenomenological
consequence of the perturbative QCD corrections and the subleading power contributions.
Interestingly, the observed strong cancellation between the two power suppressed mechanisms leads to the
insignificant correction to the leading power contribution (almost) in the whole $Q^2$ region accessible
at the current experiments. In addition, we paid a particular attention to the model dependence of the theory
predictions for $F_{\gamma^{\ast} \gamma \to \pi^0} (Q^2)$ on the twist-two pion DA.
Both the  holographic and KMOW models turned out to balance the BaBar and Belle data reasonably well at high $Q^2$,
despite the visible discrepancy at low $Q^2$ which could be compensated by the unaccounted subleading power corrections
of both perturbative and nonperturbative origins. It was also demonstrated that the end-point behaviour of the pion DA
cannot be extracted by matching the theory predictions for the pion-photon form factor with the experimental measurements.

Aiming at a better confrontation with the BaBar and Belle data, further improvements of our calculations can be made
by first carrying out the perturbative correction to the twist-4  contribution from both
the two-particle and three-particle pion DAs, which is also of conceptual interest in the framework of
perturbative QCD factorization, and then by evaluating the high twist contributions from the photon DAs
with the LCSR approach. Phenomenological applications of the techniques discussed in this work can be also pursued
in the context of the $\gamma^{\ast} \gamma \to ( \eta^{(\prime)} \,, \eta_c )$ transition form factors
\cite{Agaev:2014wna,Lees:2010de} for understanding the quark-gluon structure of eta mesons and heavy quarkonium states,
the radiative leptonic $B$-meson decays for the determination of
the inverse moment $\lambda_B$, the radiative penguin decays of $B$-mesons for the precision test of
the quark-flavour structure of the Standard Model,  and the radiative heavy-hadron decays for constraining
the magnetic susceptibility of the quark condensate \cite{Rohrwild:2007yt}.
To conclude, the  anatomy of the subleading power contributions for the exclusive hadronic reactions
is of high interest for understanding the general structures of the large momentum/mass expansion in QCD
and for hunting new physics in the quark-flavour sector as indicated by the various flavour ``anomalies"
observed at the ongoing experiments.

\subsection*{Acknowledgements}

We are grateful to Martin Beneke and Vladimir Braun for illuminating discussions
and to  Vladimir Braun for many valuable comments on the manuscript.
Y.M.W acknowledges support from the National Youth Thousand Talents Program,
the Youth Hundred Academic Leaders Program of Nankai University, and the NSFC with Grant No. 11675082.
The work of Y.L.S is supported by Natural Science Foundation of Shandong Province,
China under Grant No. ZR2015AQ006.


\appendix


\section{Two-loop evolution functions}
\label{app:two-loop evolutions}

\subsection{RG evolution of the twist-2 pion DA at two loops}

We first collect the manifest expressions of the RG functions $E_{V, n}^{\rm NLO}$ and $d_{V, n}^{k}$
appeared in  the two-loop evolution matrix of the twist-2 pion DA,  following closely \cite{Agaev:2010aq}.
Our conventions for  the QCD beta-function and the anomalous dimensions of  the local
conformal  operator \cite{Mueller:1993hg}
\begin{eqnarray}
O_{V, k}(\mu) = (i \, \bar n \cdot \partial)^k \, \bar q (0) \not \!  \bar n  \, \gamma_5 \,\,
C_k^{3/2} \left (\bar n \cdot  \buildrel\leftrightarrow\over D / \bar n \cdot \partial \right ) \,\, q(0)
\end{eqnarray}
are given by
\begin{eqnarray}
\mu {d \alpha_s(\mu) \over d \mu} &=& \beta(\alpha_s)= - 2\, \alpha_s \, \sum_{n=0} \, \beta_n \,
\, \left ( {\alpha_s \over 4 \pi} \right )^{n+1}  \,,  \\
\gamma_{V, n}(\alpha_s) &=&
- \sum_{n=0} \, \gamma_{V, n}^{(0)} \, \left ( {\alpha_s \over 4 \pi} \right )^{n+1}   \,.
\end{eqnarray}
The first three perturbative coefficients of $\beta_n$ are
\begin{eqnarray}
\beta_0 = 11 - {2 \, N_f \over 3}  \,, \qquad
\beta_1 = 102 - {38 \, N_f \over 3} \,,   \qquad
\beta_2 = {2857 \over 2} - {5033\, N_f \over 18} +
 {325\, N_f^2 \over 54} \,,
\end{eqnarray}
and the well-known LO anomalous dimension $\gamma_{V, n}^{(0)}$ reads
\begin{eqnarray}
\gamma_{V, n}^{(0)} = 2\, C_F \, \left ( 1 - {2 \over (n+1)(n+2)}
+ 4 \, \sum_{k=2}^{n+1} \, {1 \over k}  \right ) \,.
\end{eqnarray}
The NLO anomalous dimension $\gamma_{V, n}^{(1)}$ can be obtained from  the convolution integral
\begin{eqnarray}
\gamma_{V, n}^{(1)} = - {8 (2 n+ 3)\over (n+1)(n+2)} \,
\int_0^1 d x \, \int_0^1 d y \, \left [V_1(x, y) \right ]_{+}  \, y \,\bar y  \, \left [ C_n^{3/2}(2 x-1) \right ]^2 \,.
\label{convolution integral for gamma_V_NLO}
\end{eqnarray}
Making use of  the harmonic sums \cite{Floratos:1977au,GonzalezArroyo:1979df}
\begin{eqnarray}
S_l(n) = \sum_{k=1}^{n}  \, {1 \over k^l} \,, \qquad
S_l^{\prime}(n) = 2^{l-1} \, \sum_{k=1}^{n}  \, \left [1+(-1)^k \right ]\, {1 \over k^l} \,, \qquad
\tilde{S}(n) = \sum_{k=1}^{n}  \,  { (-1)^k \over k^2} \, S_1(k) \,,
\end{eqnarray}
the above-mentioned  integral (\ref{convolution integral for gamma_V_NLO}) can be further computed as \cite{Belitsky:2005qn}
\begin{eqnarray}
\gamma_{V, n}^{(1)} &=& 4 \, \left ( C_F^2 - {1 \over 2} \, C_F \, C_A \right ) \,
\bigg \{ {4 \, (2 n+3) \over (n+1)^2 \, (n+2)^2} \, S_1(n+1)
- 2 \, {3 \, n^3 + 10 \, n^2 + 11 \, n +3 \over (n+1)^3 \, (n+2)^3 }  \nonumber \\
&& + \, 4 \, \left ( 2 \, S_1(n+1) - {1 \over (n+1) (n+2)} \right ) \,
\left ( S_2(n+1) - S_2^{\prime}(n+1) \right ) + \, 16 \, \tilde{S}(n+1) \nonumber \\
&&   + \, 6\, S_2(n+1)  - {3 \over 4} - 2\, S_3^{\prime}(n+1)
- 4\, (-1)^{n+1} \, {2 \, n^2 + 6\, n + 5   \over (n+1)^3 \, (n+2)^3} \bigg \}  \nonumber \\
&& + \, 4 \, C_F\, C_A \, \bigg \{ S_1(n+1) \, \left ( {134 \over 9} + { 2 \, (2 \, n+3) \over (n+1)^2 \, (n+2)^2} \right )
- 4\, S_1(n+1) \, S_2(n+1) \, \nonumber \\
&& + \, S_2(n+1) \, \left ( - {13 \over 3} + {2 \over (n+1) (n+2) } \right )  - {43 \over 24} \nonumber \\
&& - \, {1 \over 9}   \, {151 \, n^4 + 867 \, n^3 + 1792 \, n^2 + 1590 \, n + 523  \over (n+1)^3 \, (n+2)^3 }  \bigg \} \nonumber \\
&& + \, 2 \, C_F \, N_f  \, \bigg \{ -{40 \over 9}  \, S_1(n+1)  + {8 \over 3} \,  S_2(n+1)
+ {1 \over 3}  + {4 \over 9} \, {11 \, n^2 + 27 \, n + 13  \over (n+1)^2 \, (n+2)^2 } \bigg \} \,.
\end{eqnarray}
According the master solutions displayed in (53) and (54) of \cite{Mueller:1993hg} and comparing with
(\ref{NLL evolution of pion moments}), we can readily find that \cite{Agaev:2010aq}
\begin{eqnarray}
 E_{V, n}^{\rm LO}(\mu, \mu_0) &=&
 \left ( {\alpha_s(\mu) \over \alpha_s(\mu_0)} \right )^{\gamma_{V, n}^{(0)}/(2 \, \beta_0)}\,, \nonumber \\
E_{V, n}^{\rm NLO}(\mu, \mu_0) &=& E_{V, n}^{\rm LO}(\mu, \mu_0) \,
\left \{ 1 + {\alpha_s(\mu) - \alpha_s(\mu_0) \over 8 \, \pi} \,
{\gamma_{V, n}^{(0)} \over \beta_0} \, \left ( {\gamma_{V, n}^{(1)}  \over \gamma_{V, n}^{(0)} }
- {\beta_1 \over \beta_0}  \right ) \right \}   \,,
\end{eqnarray}
and the off-diagonal evolution coefficient $d_{V, n}^{k}$ reads
\begin{eqnarray}
d_{V, n}^{k} = {M_{V, n}^k \over \gamma_{V, n}^{(0)} - \gamma_{V, k}^{(0)} - 2\, \beta_0}  \,
\left [ 1 -  \left ( {\alpha_s(\mu) \over \alpha_s(\mu_0)} \right )
^{ \left (\gamma_{V, n}^{(0)}  -\gamma_{V, n}^{(0)} - 2\, \beta_0 \right ) /(2 \, \beta_0)}  \right ]  \,.
\end{eqnarray}
The matrix element $M_{V, n}^k$ is given by
\begin{eqnarray}
M_{V, n}^k &=& {(k+1) \, (k+2) \, (2 \, n+3) \over (n+1) (n+2) } \,
\left [ \gamma_{V, n}^{(0)} - \gamma_{V, k}^{(0)}   \right ] \,
 \bigg \{ {  8 \, C_F \, A_n^k - \gamma_{V, k}^{(0)} - 2 \, \beta_0 \over  (n-k) (n+k+3)}   \nonumber \\
&& + \, 4 \, C_F \, {A_n^k - \psi(n+2) + \psi(1)  \over (k+1)(k+2)} \bigg \}  \,,
\end{eqnarray}
with
\begin{eqnarray}
A_n^k &=& \psi \left ({n+k+4 \over 2} \right ) - \psi \left ({n - k \over 2} \right )
+ 2 \, \psi (n-k) - \psi (n+2)  - \psi (1)  \,, \nonumber  \\
\psi (z) &=&  d \ln \, \Gamma(z) / d z \,.
\end{eqnarray}

\subsection{RG evolution of the twist-2 photon DA at two loops}

Along the lines of the discussion for the  pion DA, we first need the  anomalous dimensions
of the following conformal operator
\begin{eqnarray}
O_{T, k}^{\nu}(\mu) = (i \, n \cdot \partial)^k \, \bar q (0) \not \!  n  \, \gamma^{\nu, \perp} \,\,
C_k^{3/2} \left (n \cdot \buildrel\leftrightarrow\over D/n \cdot \partial \right ) \,\, q(0) \,,
\end{eqnarray}
which can be perturbatively expanded in QCD
\begin{eqnarray}
\gamma_{T, n}(\alpha_s)=
- \sum_{n=0} \, \gamma_{T, n}^{(0)} \, \left ( {\alpha_s \over 4 \pi} \right )^{n+1}   \,.
\end{eqnarray}
The one-loop anomalous dimension $\gamma_{T, n}^{(0)}$ is given by \cite{Lepage:1979zb,Shifman:1980dk}
\begin{eqnarray}
\gamma_{T, n}^{(0)} = 2\, C_F \, \left ( 1  + 4 \, \sum_{k=2}^{n+1} \, {1 \over k}  \right ) \,,
\end{eqnarray}
and the  NLO anomalous dimension $\gamma_{T, n}^{(1)}$ can be extracted from
the two-loop splitting function  for the twist-2 transversity distribution
in deep-inelastic scattering (DIS) \cite{Belitsky:2005qn,Vogelsang:1997ak,Hayashigaki:1997dn}
\begin{eqnarray}
\gamma_{T, n}^{(1)} &=&  4\, C_F^2 \,  \left [ - {1 \over 4} - 2 \, S_1(n+1) + S_2(n+1) \right ]
+ {16 \over 9} \, N_f \, C_F \, \left [ {3 \over 8} - 5 \, S_1(n+1) \, + 3 \, S_2(n+1) \right ] \nonumber \\
&&  + \, C_A \, C_F  \left [ - {20 \over 3} + {572 \over 9} \, S_1(n+1) - {58 \over 3} \, S_2(n+1)
- 16 \, S_1(n+1) \, S_2(n+1) \right ] \nonumber \\
&& - 8 \, C_F \, \left (C_F - {1 \over 2} \, C_A \right )  \,
\bigg \{ {1 \over 4} + {1 + (-1)^n \over (n+1)(n+2)} - {5 \over 2} \, S_2(n+1)
+ S_3^{\prime}(n+1) \nonumber \\
&& - 8 \, \tilde{S}(n+1) - S_1(n+1) \left [ 1 +  4 \, S_2(n+1)  - 4 \, S_2^{\prime}(n+1) \right ] \bigg \} \,.
\end{eqnarray}
The manifest expressions of the RG functions $E_{T, n}^{\rm NLO}$ and $d_{T, n}^{k}$ can  be obtained from
that of $E_{V, n}^{\rm NLO}$ and $d_{V, n}^{k}$ given above with the replacement rule
$\gamma_{V, n}^{(i)} \to \gamma_{T, n}^{(i)}$ ($i=0, \, 1$) \cite{Mueller:1993hg,Koenig:2015pha}.

\section{Spectral representations}
\label{app:spectral representations}

We present the dispersion representations of convolution integrals entering the NLL QCD
factorization formula (\ref{NLO factorization of the form factor G}) in order to construct
the sum rules for the hadronic photon correction to the pion-photon form factor.
We have verified the spectral representations in what follows numerically  by checking the corresponding
dispersion integrals.
\begin{eqnarray}
&& {1 \over \pi} \, {\rm Im}_s \, \int_0^1 \, du \, {1 \over u \, r + \bar u} \,
{ u \, r - \bar u  \over u \, \bar u \, \bar r} \, \ln (u \, r + \bar u)  \, \phi_{\gamma}(u, \mu)  \nonumber \\
&& = { Q^2 \over Q^2 + s} \, \int_0^1 \, du \, \theta \left ( u - {Q^2 \over Q^2 + s} \right ) \,
\left [ { \bar u - u \over u \, \bar u }
+ 2 \, \ln \left (  {u \, s - \bar u \, Q^2   \over Q^2} \right ) \, {d \over d u}\right ] \, \phi_{\gamma}(u, \mu) \,. \\
\nonumber  \\
&& {1 \over \pi} \, {\rm Im}_s \, \int_0^1 \, du \, {1 \over u \, r + \bar u} \,
{ u \, r - \bar u  \over u \, \bar u \, \bar r} \, \ln^2 (u \, r + \bar u)  \, \phi_{\gamma}(u, \mu)  \nonumber \\
&&=  { Q^2 \over Q^2 + s} \,  \int_0^1 \, du \, \theta \left ( u - {Q^2 \over Q^2 + s} \right ) \,
\bigg \{ \left [  2 \, \ln^2 \left (  {u \, s - \bar u \, Q^2   \over Q^2} \right )
- {2 \, \pi^2 \over 3}  \,  \right ] \, {d \over d u} \nonumber \\
&& \hspace{0.5 cm} + \, 2 \, {\bar u - u \over u \, \bar u}  \, \ln \left (  {u \, s - \bar u \, Q^2   \over Q^2} \right ) \,
 \bigg \} \, \phi_{\gamma}(u, \mu)  \,. \\
\nonumber  \\
&& {1 \over \pi} \, {\rm Im}_s \, \int_0^1 \, du \, {1 \over u \, r + \bar u} \,
{ u \, r - \bar u  \over u \, \bar u \, \bar r} \, \ln (u \, r + \bar u)  \, \ln r  \, \phi_{\gamma}(u, \mu)  \nonumber \\
&& =  { Q^2 \over Q^2 + s} \,  \int_0^1 \, du \, \bigg \{  {\bar u - u \over u \, \bar u}  \,
\left [ \theta \left ( u - {Q^2 \over Q^2 + s} \right ) \, \ln \left ( {s \over Q^2} \right )
+  \ln  \bigg|  {u \, s - \bar u \, Q^2   \over Q^2}  \bigg |  \right ] \, \phi_{\gamma}(u, \mu) \nonumber \\
&& \hspace{0.5 cm}  + \left [  \ln^2  \bigg|  {u \, s - \bar u \, Q^2   \over Q^2}  \bigg |
+  \theta \left ( u - {Q^2 \over Q^2 + s} \right ) \,
\left ( 2 \,  \ln \left (  {u \, s - \bar u \, Q^2   \over Q^2} \right )  \, \ln \left ( {s \over Q^2} \right )
- \pi^2  \right ) \right ]   \nonumber \\
&&  \hspace{0.8 cm} \times  {d \over d u} \,  \phi_{\gamma}(u, \mu)   \bigg \}     \,.   \\
\nonumber  \\
&& {1 \over \pi} \, {\rm Im}_s \, \int_0^1 \, du \, {1 \over u \, r + \bar u} \,
{ r   \over \bar u \, \bar r} \,  \ln r  \, \phi_{\gamma}(u, \mu)  \nonumber \\
&& =  - {Q^2 \over Q^2 + s} \, \ln \left ( {s \over Q^2} \right ) \, \phi_{\gamma}\left({Q^2 \over Q^2 + s}, \mu \right)
-   \int_0^1 \, {du \over \bar u} \,  \left [ {Q^2 \over s + Q^2}
+ {\cal P} {\bar u \, Q^2 \over u \, s - \bar u \, Q^2} \right ] \, \phi_{\gamma}(u, \mu)    \,.  \hspace{0.5 cm}  \\
\nonumber  \\
&& {1 \over \pi} \, {\rm Im}_s \, \int_0^1 \, du \, {1 \over u \, r + \bar u} \,
{ r   \over \bar u \, \bar r} \,   \ln (u \, r + \bar u) \ln r  \, \phi_{\gamma}(u, \mu)  \nonumber \\
&& = {Q^2 \over Q^2+s} \,  \int_0^1 \, du \, \bigg \{ - {1 \over \bar u}  \,
\left [ \theta \left ( u - {Q^2 \over Q^2 + s} \right ) \, \ln \left ( {s \over Q^2} \right )
+  \ln  \bigg|  {u \, s - \bar u \, Q^2   \over Q^2}  \bigg |  \right ] \, \phi_{\gamma}(u, \mu)  \nonumber \\
&& \hspace{0.5 cm}  + \,  {1 \over 2}  \, \left [  \ln^2  \bigg|  {u \, s - \bar u \, Q^2   \over Q^2}  \bigg |
+  \theta \left ( u - {Q^2 \over Q^2 + s} \right ) \,
\left ( 2 \,  \ln \left (  {u \, s - \bar u \, Q^2   \over Q^2} \right )  \, \ln \left ( {s \over Q^2} \right )
- \pi^2  \right ) \right ]   \nonumber \\
&&  \hspace{0.8 cm} \times  {d \over d u} \,  \phi_{\gamma}(u, \mu)   \bigg \}     \,. \\
\nonumber  \\
&& {1 \over \pi} \, {\rm Im}_s \, \int_0^1 \, du \, {1 \over u \, r + \bar u} \,
{ r   \over \bar u \, \bar r} \,   \ln^2 r  \, \phi_{\gamma}(u, \mu)  \nonumber \\
&& =   - {Q^2 \over Q^2+s} \, \left[\ln^2 \left ( {s \over Q^2} \right )  - \pi^2  \right ]  \,
\phi_{\gamma}\left({Q^2 \over Q^2 + s}, \mu \right)  \nonumber \\
&& \hspace{0.5 cm} - 2\, \ln \left ( {s \over Q^2} \right ) \, \int_0^1 \, { d u \over \bar u} \,
\left [ {Q^2 \over Q^2+s}  + {\cal P} {\bar u \, Q^2 \over u \, s - \bar u \, Q^2}  \right ] \,   \phi_{\gamma}(u, \mu) \,. \\
\nonumber  \\
&& {1 \over \pi} \, {\rm Im}_s \, \int_0^1 \, du \, {1 \over u \, r + \bar u} \,
{ 1   \over  u \, \bar r} \,   \left ( \ln r + 3 \right ) \,
\ln \left (\bar u + u \, r   \right )   \, \phi_{\gamma}(u, \mu) \nonumber \\
&& = {Q^2 \over Q^2+s} \, \int_0^1 \, du \,
\bigg \{ - { \phi_{\gamma}(u, \mu) \over u} \, \left [ \theta \left ( u - {Q^2 \over Q^2 + s} \right ) \,
\left (\ln \left ( {s \over Q^2} \right )  + 3 \right )
+  \ln  \bigg|  {u \, s - \bar u \, Q^2   \over Q^2}  \bigg |   \right ] \nonumber \\
&& \hspace{0.5 cm} - {1 \over 2} \, \left [\ln^2  \bigg|  {u \, s - \bar u \, Q^2   \over Q^2}  \bigg |
+  \theta \left ( u - {Q^2 \over Q^2 + s} \right ) \,
\left (  \left ( 2 \,  \ln \left ( {s \over Q^2} \right )  + 3 \right )\,
\ln \left (  {u \, s - \bar u \, Q^2   \over Q^2}  \right )   - \pi^2 \right ) \, \right ] \nonumber \\
&& \hspace{0.8 cm} \times {d \over d u}  \,  \phi_{\gamma}(u, \mu)  \bigg \}      \,.
\end{eqnarray}
Here,  the parameter $p^2$ in the definition of $r$ should be obviously understood as $s$ in the above convolution integrals
and ${\cal P}$ represents the principle-value prescription.



\begin{thebibliography}{99}



\bibitem{Lepage:1980fj}
  G.~P.~Lepage and S.~J.~Brodsky,
  Phys.\ Rev.\ D {\bf 22} (1980) 2157.





\bibitem{Efremov:1979qk}
  A.~V.~Efremov and A.~V.~Radyushkin,
  Phys.\ Lett.\  {\bf 94B} (1980) 245.





\bibitem{Duncan:1979ny}
  A.~Duncan and A.~H.~M\"{u}eller,
  Phys.\ Lett.\  {\bf 90B} (1980) 159.





\bibitem{Rothstein:2003wh}
  I.~Z.~Rothstein,
  Phys.\ Rev.\ D {\bf 70} (2004) 054024
  [hep-ph/0301240].





\bibitem{delAguila:1981nk}
  F.~del Aguila and M.~K.~Chase,
  Nucl.\ Phys.\ B {\bf 193} (1981) 517.




\bibitem{Braaten:1982yp}
  E.~Braaten,
  Phys.\ Rev.\ D {\bf 28} (1983) 524.




\bibitem{Kadantseva:1985kb}
  E.~P.~Kadantseva, S.~V.~Mikhailov and A.~V.~Radyushkin,
  Yad.\ Fiz.\  {\bf 44} (1986) 507
   [Sov.\ J.\ Nucl.\ Phys.\  {\bf 44} (1986) 326].




\bibitem{Melic:2002ij}
  B.~Melic, D.~M\"{u}eller and K.~Passek-Kumericki,
  Phys.\ Rev.\ D {\bf 68} (2003) 014013
  [hep-ph/0212346].




\bibitem{Bonneau:1990xu}
  G.~Bonneau,
  Int.\ J.\ Mod.\ Phys.\ A {\bf 5} (1990) 3831.





\bibitem{Collins:1984xc}
  J.~C.~Collins,
  ``{\it Renormalization : An Introduction to Renormalization, The Renormalization Group, and the Operator Product Expansion},''
  Cambridge University Press, 1984.





\bibitem{Larin:1993tq}
  S.~A.~Larin,
  Phys.\ Lett.\ B {\bf 303} (1993) 113
  [hep-ph/9302240].



\bibitem{Martin:1999cc}
  C.~P.~Martin and D.~Sanchez-Ruiz,
  Nucl.\ Phys.\ B {\bf 572} (2000) 387
  [hep-th/9905076].



\bibitem{Jegerlehner:2000dz}
  F.~Jegerlehner,
  Eur.\ Phys.\ J.\ C {\bf 18} (2001) 673
  [hep-th/0005255].




\bibitem{Moch:2015usa}
  S.~Moch, J.~A.~M.~Vermaseren and A.~Vogt,
  Phys.\ Lett.\ B {\bf 748} (2015) 432
  [arXiv:1506.04517 [hep-ph]].





\bibitem{Gutierrez-Reyes:2017glx}
  D.~Guti¨¦rrez-Reyes, I.~Scimemi and A.~A.~Vladimirov,
  Phys.\ Lett.\ B {\bf 769} (2017) 84
  [arXiv:1702.06558 [hep-ph]].




\bibitem{Beneke:2004rc}
  M.~Beneke, Y.~Kiyo and D.~S.~Yang,
  Nucl.\ Phys.\ B {\bf 692} (2004) 232
  [hep-ph/0402241].




\bibitem{Beneke:2005gs}
  M.~Beneke and D.~Yang,
  Nucl.\ Phys.\ B {\bf 736} (2006) 34
  [hep-ph/0508250].





\bibitem{Beneke:2005vv}
  M.~Beneke and S.~Jager,
  Nucl.\ Phys.\ B {\bf 751} (2006) 160
  [hep-ph/0512351].




\bibitem{Dugan:1990df}
  M.~J.~Dugan and B.~Grinstein,
  Phys.\ Lett.\ B {\bf 256} (1991) 239.





\bibitem{Herrlich:1994kh}
  S.~Herrlich and U.~Nierste,
  Nucl.\ Phys.\ B {\bf 455} (1995) 39
  [hep-ph/9412375].




\bibitem{Aubert:2009mc}
  B.~Aubert {\it et al.} [BaBar Collaboration],
  Phys.\ Rev.\ D {\bf 80} (2009) 052002
  [arXiv:0905.4778 [hep-ex]].





\bibitem{Agaev:2010aq}
  S.~S.~Agaev, V.~M.~Braun, N.~Offen and F.~A.~Porkert,
  Phys.\ Rev.\ D {\bf 83} (2011) 054020
  [arXiv:1012.4671 [hep-ph]].




\bibitem{Agaev:2012tm}
  S.~S.~Agaev, V.~M.~Braun, N.~Offen and F.~A.~Porkert,
  Phys.\ Rev.\ D {\bf 86} (2012) 077504
  [arXiv:1206.3968 [hep-ph]].



\bibitem{Kroll:2010bf}
  P.~Kroll,
  Eur.\ Phys.\ J.\ C {\bf 71} (2011) 1623
  [arXiv:1012.3542 [hep-ph]].





\bibitem{Li:2013xna}
  H.~N.~Li, Y.~L.~Shen and Y.~M.~Wang,
  JHEP {\bf 1401} (2014) 004
  [arXiv:1310.3672 [hep-ph]].





\bibitem{Khodjamirian:1997tk}
  A.~Khodjamirian,
  Eur.\ Phys.\ J.\ C {\bf 6} (1999) 477
  [hep-ph/9712451].




\bibitem{Stefanis:2012yw}
  N.~G.~Stefanis, A.~P.~Bakulev, S.~V.~Mikhailov and A.~V.~Pimikov,
  Phys.\ Rev.\ D {\bf 87} (2013)  094025
  [arXiv:1202.1781 [hep-ph]].




\bibitem{Bakulev:2011rp}
  A.~P.~Bakulev, S.~V.~Mikhailov, A.~V.~Pimikov and N.~G.~Stefanis,
  Phys.\ Rev.\ D {\bf 84} (2011) 034014
  [arXiv:1105.2753 [hep-ph]].




\bibitem{Bakulev:2012nh}
  A.~P.~Bakulev, S.~V.~Mikhailov, A.~V.~Pimikov and N.~G.~Stefanis,
  Phys.\ Rev.\ D {\bf 86} (2012) 031501
  [arXiv:1205.3770 [hep-ph]].




\bibitem{Mikhailov:2016klg}
  S.~V.~Mikhailov, A.~V.~Pimikov and N.~G.~Stefanis,
  Phys.\ Rev.\ D {\bf 93} (2016)   114018
  [arXiv:1604.06391 [hep-ph]].





\bibitem{Braun:2012kp}
  V.~M.~Braun and A.~Khodjamirian,
  Phys.\ Lett.\ B {\bf 718} (2013) 1014
  [arXiv:1210.4453 [hep-ph]].



\bibitem{Wang:2016qii}
  Y.~M.~Wang,
  JHEP {\bf 1609} (2016) 159
  [arXiv:1606.03080 [hep-ph]].




\bibitem{Agaev:2014wna}
  S.~S.~Agaev, V.~M.~Braun, N.~Offen, F.~A.~Porkert and A.~Sch\"{a}fer,
  Phys.\ Rev.\ D {\bf 90} (2014) 074019
  [arXiv:1409.4311 [hep-ph]].





\bibitem{Braun:2016tsk}
  V.~M.~Braun, N.~Kivel, M.~Strohmaier and A.~A.~Vladimirov,
  JHEP {\bf 1606} (2016) 039
  [arXiv:1603.09154 [hep-ph]].



\bibitem{Ball:2002ps}
  P.~Ball, V.~M.~Braun and N.~Kivel,
  Nucl.\ Phys.\ B {\bf 649} (2003) 263
  [hep-ph/0207307].




\bibitem{Li:1992nu}
  H.~n.~Li and G.~F.~Sterman,
  Nucl.\ Phys.\ B {\bf 381} (1992) 129.





\bibitem{Nandi:2007qx}
  S.~Nandi and H.~n.~Li,
  Phys.\ Rev.\ D {\bf 76} (2007) 034008
  [arXiv:0704.3790 [hep-ph]].





\bibitem{Musatov:1997pu}
  I.~V.~Musatov and A.~V.~Radyushkin,
  Phys.\ Rev.\ D {\bf 56} (1997) 2713
  [hep-ph/9702443].




\bibitem{Wu:2010zc}
  X.~G.~Wu and T.~Huang,
  Phys.\ Rev.\ D {\bf 82} (2010) 034024
  [arXiv:1005.3359 [hep-ph]].







\bibitem{Chen:2011pn}
  Y.~C.~Chen and H.~n.~Li,
  Phys.\ Rev.\ D {\bf 84} (2011) 034018
  [arXiv:1104.5398 [hep-ph]].




\bibitem{Li:2014xda}
  H.~n.~Li and Y.~M.~Wang,
  JHEP {\bf 1506} (2015) 013
  [arXiv:1410.7274 [hep-ph]].


\bibitem{He:2006ud}
  X.~G.~He, T.~Li, X.~Q.~Li and Y.~M.~Wang,
  Phys.\ Rev.\ D {\bf 74} (2006) 034026
  [hep-ph/0606025].




\bibitem{Lu:2009cm}
  C.~D.~L\"{u}, Y.~M.~Wang, H.~Zou, A.~Ali and G.~Kramer,
  Phys.\ Rev.\ D {\bf 80} (2009) 034011
  [arXiv:0906.1479 [hep-ph]].




\bibitem{Li:2010nn}
  H.~n.~Li, Y.~L.~Shen, Y.~M.~Wang and H.~Zou,
  Phys.\ Rev.\ D {\bf 83} (2011) 054029
  [arXiv:1012.4098 [hep-ph]].




\bibitem{Li:2012nk}
  H.~n.~Li, Y.~L.~Shen and Y.~M.~Wang,
  Phys.\ Rev.\ D {\bf 85} (2012) 074004
  [arXiv:1201.5066 [hep-ph]].




\bibitem{Li:2012md}
  H.~N.~Li, Y.~L.~Shen and Y.~M.~Wang,
  JHEP {\bf 1302} (2013) 008
  [arXiv:1210.2978 [hep-ph]].





\bibitem{Uehara:2012ag}
  S.~Uehara {\it et al.} [Belle Collaboration],
  Phys.\ Rev.\ D {\bf 86} (2012) 092007
  [arXiv:1205.3249 [hep-ex]].






\bibitem{Masjuan:2012wy}
  P.~Masjuan,
  Phys.\ Rev.\ D {\bf 86} (2012) 094021
  [arXiv:1206.2549 [hep-ph]].






\bibitem{Hoferichter:2014vra}
  M.~Hoferichter, B.~Kubis, S.~Leupold, F.~Niecknig and S.~P.~Schneider,
  Eur.\ Phys.\ J.\ C {\bf 74} (2014) 3180
  [arXiv:1410.4691 [hep-ph]].




\bibitem{Gerardin:2016cqj}
  A.~G¨¦rardin, H.~B.~Meyer and A.~Nyffeler,
  Phys.\ Rev.\ D {\bf 94} (2016)   074507
  [arXiv:1607.08174 [hep-lat]].



\bibitem{Radyushkin:2009zg}
  A.~V.~Radyushkin,
  Phys.\ Rev.\ D {\bf 80} (2009) 094009
  [arXiv:0906.0323 [hep-ph]].




\bibitem{Polyakov:2009je}
  M.~V.~Polyakov,
  JETP Lett.\  {\bf 90} (2009) 228
  [arXiv:0906.0538 [hep-ph]].







\bibitem{Matiounine:1998re}
  Y.~Matiounine, J.~Smith and W.~L.~van Neerven,
  Phys.\ Rev.\ D {\bf 58} (1998) 076002
  [hep-ph/9803439].






\bibitem{Ravindran:2003gi}
  V.~Ravindran, J.~Smith and W.~L.~van Neerven,
  Nucl.\ Phys.\ B {\bf 682} (2004) 421
  [hep-ph/0311304].




\bibitem{Beneke:1997zp}
  M.~Beneke and V.~A.~Smirnov,
  Nucl.\ Phys.\ B {\bf 522} (1998) 321
  [hep-ph/9711391].




\bibitem{Melic:2001wb}
  B.~Melic, B.~Nizic and K.~Passek,
  Phys.\ Rev.\ D {\bf 65} (2002) 053020
  [hep-ph/0107295].




\bibitem{Sarmadi:1982yg}
  M.~H.~Sarmadi,
  Phys.\ Lett.\  {\bf 143B} (1984) 471.





\bibitem{Dittes:1983dy}
  F.~M.~Dittes and A.~V.~Radyushkin,
  Phys.\ Lett.\  {\bf 134B} (1984) 359.




\bibitem{Katz:1984gf}
  G.~R.~Katz,
  Phys.\ Rev.\ D {\bf 31} (1985) 652.





\bibitem{Mikhailov:1984ii}
  S.~V.~Mikhailov and A.~V.~Radyushkin,
  Nucl.\ Phys.\ B {\bf 254} (1985) 89.





\bibitem{Belitsky:1999gu}
  A.~V.~Belitsky, D.~M\"{u}ller and A.~Freund,
  Phys.\ Lett.\ B {\bf 461} (1999) 270
  [hep-ph/9904477].




\bibitem{Mikhailov:1985cm}
  S.~V.~Mikhailov and A.~V.~Radyushkin,
  Nucl.\ Phys.\ B {\bf 273} (1986) 297.




\bibitem{Mueller:1993hg}
  D.~M\"{u}ller,
  Phys.\ Rev.\ D {\bf 49} (1994) 2525.




\bibitem{Mueller:1994cn}
  D.~M\"{u}ller,
  Phys.\ Rev.\ D {\bf 51} (1995) 3855
  [hep-ph/9411338].






\bibitem{Grossmann:2015lea}
  Y.~Grossman, M.~K\"{o}nig and M.~Neubert,
  JHEP {\bf 1504} (2015) 101
  [arXiv:1501.06569 [hep-ph]].





\bibitem{Ioffe:1983ju}
  B.~L.~Ioffe and A.~V.~Smilga,
  Nucl.\ Phys.\ B {\bf 232} (1984) 109.




\bibitem{Koenig:2015pha}
  M.~K\"{o}nig and M.~Neubert,
  JHEP {\bf 1508} (2015) 012
  [arXiv:1505.03870 [hep-ph]].




\bibitem{Shifman:1980dk}
  M.~A.~Shifman and M.~I.~Vysotsky,
  Nucl.\ Phys.\ B {\bf 186} (1981) 475.





\bibitem{Wang:2013ywc}
  X.~P.~Wang and D.~S.~Yang,
  JHEP {\bf 1406} (2014) 121
  [arXiv:1401.0122 [hep-ph]].



\bibitem{Lepage:1979zb}
  G.~P.~Lepage and S.~J.~Brodsky,
  Phys.\ Lett.\  {\bf 87B} (1979) 359.


\bibitem{Chetyrkin:1997dh}
  K.~G.~Chetyrkin,
  Phys.\ Lett.\ B {\bf 404} (1997) 161
  [hep-ph/9703278].



\bibitem{Vermaseren:1997fq}
  J.~A.~M.~Vermaseren, S.~A.~Larin and T.~van Ritbergen,
  Phys.\ Lett.\ B {\bf 405} (1997) 327
  [hep-ph/9703284].



\bibitem{Wang:2015ndk}
  Y.~M.~Wang and Y.~L.~Shen,
  JHEP {\bf 1602} (2016) 179
  [arXiv:1511.09036 [hep-ph]].




\bibitem{Mikhailov:2008my}
  S.~V.~Mikhailov and A.~A.~Vladimirov,
  Phys.\ Lett.\ B {\bf 671} (2009) 111
  [arXiv:0810.1647 [hep-ph]].



\bibitem{Belitsky:2000yn}
  A.~V.~Belitsky, A.~Freund and D.~M\"{u}ller,
  Phys.\ Lett.\ B {\bf 493} (2000) 341
  [hep-ph/0008005].




\bibitem{Braun:2017cih}
  V.~M.~Braun, A.~N.~Manashov, S.~Moch and M.~Strohmaier,
  JHEP {\bf 1706} (2017) 037
  [arXiv:1703.09532 [hep-ph]].






\bibitem{Chernyak:1981zz}
  V.~L.~Chernyak and A.~R.~Zhitnitsky,
  Nucl.\ Phys.\ B {\bf 201} (1982) 492
   Erratum: [Nucl.\ Phys.\ B {\bf 214} (1983) 547].





\bibitem{Bakulev:2001pa}
  A.~P.~Bakulev, S.~V.~Mikhailov and N.~G.~Stefanis,
  Phys.\ Lett.\ B {\bf 508} (2001) 279
   Erratum: [Phys.\ Lett.\ B {\bf 590} (2004) 309]
  [hep-ph/0103119].





\bibitem{Chernyak:2006ms}
  V.~L.~Chernyak,
  Nucl.\ Phys.\ Proc.\ Suppl.\  {\bf 162} (2006) 161
  [hep-ph/0605327].



\bibitem{Khodjamirian:2011ub}
  A.~Khodjamirian, T.~Mannel, N.~Offen and Y.-M.~Wang,
  Phys.\ Rev.\ D {\bf 83} (2011) 094031
  [arXiv:1103.2655 [hep-ph]].



\bibitem{Brodsky:2007hb}
  S.~J.~Brodsky and G.~F.~de Teramond,
  Phys.\ Rev.\ D {\bf 77} (2008) 056007
  [arXiv:0707.3859 [hep-ph]].






\bibitem{Cloet:2013tta}
  I.~C.~Clo\"{e}t, L.~Chang, C.~D.~Roberts, S.~M.~Schmidt and P.~C.~Tandy,
  Phys.\ Rev.\ Lett.\  {\bf 111} (2013) 092001
  [arXiv:1306.2645 [nucl-th]].





\bibitem{Braun:2015axa}
  V.~M.~Braun, S.~Collins, M.~G\"{o}ckeler, P.~P\'{e}rez-Rubio, A.~Sch\"{a}fer, R.~W.~Schiel and A.~Sternbeck,
  Phys.\ Rev.\ D {\bf 92} (2015)  014504
  [arXiv:1503.03656 [hep-lat]].




\bibitem{Novikov:1983jt}
  V.~A.~Novikov, M.~A.~Shifman, A.~I.~Vainshtein, M.~B.~Voloshin and V.~I.~Zakharov,
  Nucl.\ Phys.\ B {\bf 237} (1984) 525.




\bibitem{Ball:2006wn}
  P.~Ball, V.~M.~Braun and A.~Lenz,
  JHEP {\bf 0605} (2006) 004
  [hep-ph/0603063].




\bibitem{Rohrwild:2007yt}
  J.~Rohrwild,
  JHEP {\bf 0709} (2007) 073
  [arXiv:0708.1405 [hep-ph]].




\bibitem{Vainshtein:2002nv}
  A.~Vainshtein,
  Phys.\ Lett.\ B {\bf 569} (2003) 187
  [hep-ph/0212231].



\bibitem{Wang:2015vgv}
  Y.~M.~Wang and Y.~L.~Shen,
  Nucl.\ Phys.\ B {\bf 898} (2015) 563
  [arXiv:1506.00667 [hep-ph]].




\bibitem{Wang:2017jow}
  Y.~M.~Wang, Y.~B.~Wei, Y.~L.~Shen and C.~D.~L\"{u},
  arXiv:1701.06810 [hep-ph].






\bibitem{Khodjamirian:2006st}
  A.~Khodjamirian, T.~Mannel and N.~Offen,
  Phys.\ Rev.\ D {\bf 75} (2007) 054013
  [hep-ph/0611193].




\bibitem{Gronberg:1997fj}
  J.~Gronberg {\it et al.} [CLEO Collaboration],
  Phys.\ Rev.\ D {\bf 57} (1998) 33
  [hep-ex/9707031].







\bibitem{Bali:2017ude}
  G.~S.~Bali {\it et al.} [RQCD Collaboration],
  arXiv:1705.10236 [hep-lat].





\bibitem{Unverzagt:2012zz}
  M.~Unverzagt,
  J.\ Phys.\ Conf.\ Ser.\  {\bf 349} (2012) 012015.






\bibitem{Lees:2010de}
  J.~P.~Lees {\it et al.} [BaBar Collaboration],
  Phys.\ Rev.\ D {\bf 81} (2010) 052010
  [arXiv:1002.3000 [hep-ex]].






\bibitem{Floratos:1977au}
  E.~G.~Floratos, D.~A.~Ross and C.~T.~Sachrajda,
  Nucl.\ Phys.\ B {\bf 129} (1977) 66
   Erratum: [Nucl.\ Phys.\ B {\bf 139} (1978) 545].




\bibitem{GonzalezArroyo:1979df}
  A.~Gonzalez-Arroyo, C.~Lopez and F.~J.~Yndurain,
  Nucl.\ Phys.\ B {\bf 153} (1979) 161.


\bibitem{Belitsky:2005qn}
  A.~V.~Belitsky and A.~V.~Radyushkin,
  Phys.\ Rept.\  {\bf 418} (2005) 1
  [hep-ph/0504030].
  
  
  

\bibitem{Vogelsang:1997ak}
  W.~Vogelsang,
  Phys.\ Rev.\ D {\bf 57} (1998) 1886
  [hep-ph/9706511].




\bibitem{Hayashigaki:1997dn}
  A.~Hayashigaki, Y.~Kanazawa and Y.~Koike,
  Phys.\ Rev.\ D {\bf 56} (1997) 7350
  [hep-ph/9707208].










\end{thebibliography}
\end{document}